\documentclass[twocolumn]{aastex62}
\usepackage{amsmath,amstext}
\usepackage[T1]{fontenc}
\usepackage{apjfonts} 
\usepackage[figure,figure*]{hypcap}
\usepackage{longtable}
\usepackage{amssymb}
\usepackage{epsfig}
\usepackage{thumbpdf}
\usepackage{epstopdf}
\usepackage{graphicx}
\usepackage{soul}
\usepackage[english]{babel}
\usepackage[utf8]{inputenc}
\usepackage{natbib}
\usepackage{xargs}                   
\newcommandx{\unsure}[2][1=]{\todo[linecolor=red,backgroundcolor=red!25,bordercolor=red,#1]{#2}} 
\newcommandx{\change}[2][1=]{\todo[linecolor=blue,backgroundcolor=blue!25,bordercolor=blue,#1]{#2}} 
\newcommandx{\info}[2][1=]{\todo[linecolor=OliveGreen,backgroundcolor=OliveGreen!25,bordercolor=OliveGreen,#1]{#2}} 
\newcommandx{\improvement}[2][1=]{\todo[linecolor=Plum,backgroundcolor=Plum!25,bordercolor=Plum,#1]{#2}}
\newcommandx{\thiswillnotshow}[2][1=]{\todo[disable,#1]{#2}}
\usepackage{lmacs}
\usepackage{savesym}
 
\def\Zsun{${\rm Z}_{\odot}$}
\def\Msun{${\rm M}_{\odot}$}

\def\EBV{E($B-V$)}
\def\Rv{$R_V$}

\shorttitle{PISCO}
\shortauthors{Galbany et al.}

\begin{document}

\title{PISCO: THE PMAS/PPAK INTEGRAL-FIELD SUPERNOVA HOSTS COMPILATION}

\correspondingauthor{Llu\'is Galbany} \email{llgalbany@pitt.edu}
\author[0000-0002-1296-6887]{L. Galbany}
\affiliation{PITT PACC, Department of Physics and Astronomy, University of Pittsburgh, Pittsburgh, PA 15260, USA.}
\author{J. P. Anderson}
\affiliation{European Southern Observatory, Alonso de Cordova 3107 Casilla 19001 $-$ Vitacura $-$ Santiago, Chile.}
\author{S. F. S\'anchez}
\affiliation{Instituto de Astronom\'ia, Universidad Nacional Aut\'onoma de M\'exico, A.P. 70-264, 04510 M\'exico, D.F., Mexico.}
\author{H. Kuncarayakti}
\affiliation{Finnish Centre for Astronomy with ESO (FINCA), University of Turku, V\"ais\"al\"antie 20, 21500 Piikki\"o, Finland.}
\affiliation{Tuorla Observatory, Department of Physics and Astronomy, University of Turku, V\"ais\"al\"antie 20, 21500 Piikki\"o, Finland.}
\author[0000-0003-1346-208X]{S. Pedraz} 
\affiliation{Centro Astron\'omico Hispano-Alem\'an (CSIC-MPG), Observatorio Astron\'omico de Calar Alto, Sierra de los Filabres - 04550 G\'ergal, Almer\'ia, Spain.}
\author{S. Gonz\'alez-Gait\'an}
\affiliation{CENTRA - Centro de Astrof\'isica e Gravitação, Instituto Superior T\'ecnico, Av. Rovisco Pais 1, 1049-001, Lisbon, Portugal.}
\author{V. Stanishev}
\affiliation{Department of Physics, Chemistry and Biology, IFM, Link\"oping University, 581 83 Link\"oping, Sweden}
\author{I. Dom\'inguez}
\affiliation{Dpto. F\'isica Te\'orica y del Cosmos, Universidad de Granada, E-18071 Granada, Spain.}
\author{M. E. Moreno-Raya}
\affiliation{Centro Astron\'omico Hispano-Alem\'an (CSIC-MPG), Observatorio Astron\'omico de Calar Alto, Sierra de los Filabres - 04550 G\'ergal, Almer\'ia, Spain.}
\author{W. M. Wood-Vasey}
\affiliation{PITT PACC, Department of Physics and Astronomy, University of Pittsburgh, Pittsburgh, PA 15260, USA.}
\author{A. M. Mour\~ao}
\affiliation{CENTRA - Centro de Astrof\'isica e Gravitação, Instituto Superior T\'ecnico, Av. Rovisco Pais 1, 1049-001, Lisbon, Portugal.}
\author{K. A. Ponder}
\affiliation{Berkeley Center for Cosmological Physics, Campbell Hall 341, University of California, Berkeley, CA 94720, USA.}
\author{C. Badenes}
\affiliation{PITT PACC, Department of Physics and Astronomy, University of Pittsburgh, Pittsburgh, PA 15260, USA.}
\author{M. Moll\'a}
\affiliation{Departamento de Investigaci\'on B\'asica, CIEMAT, Avda. Complutense 40, 28040, Madrid, Spain.}
\author{A. R. L\'opez-S\'anchez}
\affiliation{Australian Astronomical Observatory, PO Box 915, North Ryde, NSW 1670, Australia.}
\affiliation{Department of Physics and Astronomy, Macquarie University, NSW 2109, Australia.}
\author{F. F. Rosales-Ortega}
\affiliation{Instituto Nacional de Astrof\'isica, \'Optica y Electr\'onica (INAOE), 72840 Tonantzintla, Puebla, Mexico.}
\author{J. M. V\'ilchez}
\affiliation{Instituto de Astrof\'isica de Andaluc\'ia (IAA/CSIC), Glorieta de la Astronom\'ia s/n Aptdo. 3004, E-18080 Granada, Spain.}
\author[0000-0002-7077-308X]{R. Garc\'ia-Benito}               
\affiliation{Instituto de Astrof\'isica de Andaluc\'ia (IAA/CSIC), Glorieta de la Astronom\'ia s/n Aptdo. 3004, E-18080 Granada, Spain.}
\author{R. A. Marino}                   
\affiliation{ETH Z\"urich, Department of Physics, Wolfgang-Pauli-Str. 27, 8093 Z\"urich, Switzerland.}

\begin{abstract}
We present the PMAS/PPak Integral-field Supernova hosts COmpilation (PISCO) which comprises integral field spectroscopy (IFS) of 232 supernova (SN) host galaxies, that hosted 272 SNe, observed over several semesters with the 3.5m telescope at the Calar Alto Observatory (CAHA). 
PISCO is the largest collection of SN host galaxies observed with wide-field IFS, totaling 466,347 individual spectra covering a typical spatial resolution of $\sim$380 pc.
While focused studies regarding specific SN Ia-related topics will be published elsewhere, this paper aims to present the properties of the SN environments with stellar population (SP) synthesis and the gas-phase interstellar medium, providing additional results separating stripped-envelope SNe into their subtypes.
With 11,270 {\sc Hii} regions detected in all galaxies, we present for the first time an {\sc Hii} region statistical analysis, that puts {\sc Hii} regions that have hosted SNe in context with all other star forming clumps within their galaxies. 
SNe Ic are associated to more metal-rich, higher EW(H$\alpha$) and higher star formation rate (SFR) environments within their host galaxies than the mean of all HII regions detected within each host, 
on contrary SNe IIb occur at the most different environments compared to other core-collapse SNe types.
We find two clear components of young and old SP at SNe IIn locations.
We find that SNe II fast-decliners (IIL) tend to explode at locations where  
$\Sigma_{\rm SFR}$ is more intense. 
Finally, we outline how a future dedicated IFS survey of galaxies in parallel to an untargeted SN search would overcome the biases in current environmental studies.
\end{abstract}


\received{December 19, 2017}
\revised{February 1, 2018}
\accepted{February 4, 2018}


\section{Introduction}

Supernovae (SNe) are the main driver of chemical enrichment of the interstellar medium in galaxies \citep{1986A&A...154..279M,2010ApJ...714L.275M} by propelling the elements created in the interior of their progenitor stars and during the explosion \citep{1960ApJ...132..565H,1986ARA&A..24..205W}, and also one of the main producers of dust grains \citep{2001MNRAS.325..726T,2007MNRAS.378..973B}. 
However, the ultimate link between different kinds of SN explosions and their progenitor systems is far from being identified, and it is also key for our understanding of stellar evolution since various models predict different end-paths for stars in a range of masses, metallicities and ages.

Although SNe Ia are the most precise extragalactic distance indicator to date \citep{2014A&A...568A..22B,2017arXiv171000845S}, the exact understanding of the progenitor systems and explosion mechanism of SNe Ia remains elusive.
There is a wide consensus in the community that their progenitors are low mass stars (1 < M < 8 \Msun; \citealt{1980ApJ...237..111B}) that evolve to form degenerate carbon-oxygen (C/O) white dwarfs (WD) with masses in the range of 0.5 < M < 1.1 \Msun~\citep{1999ApJ...524..226D}. 
A WD in a binary system can accrete mass from its companion star and, under certain conditions, increase its mass to the point that thermonuclear reactions can ignite in its center to completely disrupt the star \citep{1960ApJ...132..565H}.
Some heterogeneities in the observed properties of SNe Ia can be attributed to differences in the nature of the companion star (another WD or non-degenerate donor star), the actual mass (below or close to the Chandrasekhar limit) of the WD at the explosion \citep{2014ARA&A..52..107M}, or in differences in how the explosion initiates and propagates (as a detonation or a deflagration).

On the other hand, single-star evolution models predict that massive stars (M $>$ 8 \Msun) form a heavy iron core 4-40 Myr after their birth, which gravitationally collapses into a neutron star or a black hole triggering the explosive ejection of the star outer envelope, and producing a core collapse supernovae (CC SN; \citealt{1979NuPhA.324..487B,1989ARA&A..27..629A}). 
Type II SNe are the most common CC SN. They retain their external H layer prior to explosion, and have also proved to be useful distance indicators (e.g., \citealt{2017MNRAS.472.4233D}).
Progenitor detections in pre-explosion images have constrained the initial mass to be between 8.5 and 16.5 \Msun, the most viable candidates being red supergiants (RSG; \citealt{2015PASA...32...16S}). 
Historically, SNe II have been divided in two families, {\it plateau} (IIP) and {\it linear} (IIL), depending on their post-maximum brightness decay rate, but current evidence have probed this separation to be superfluous since all SNII cover a continuous range of decay rates \citep{2014ApJ...786...67A,2015ApJ...799..208S, 2016AJ....151...33G}.
Type IIn SNe show narrow lines in their spectra resulting from interaction between the ejecta and circumstellar matter (CSM). Although only a few progenitor detections have been reported \citep{2009Natur.458..865G}, their progenitors could be less massive than the normal SN II \citep{2012MNRAS.424.1372A,2014MNRAS.441.2230H}.

A small fraction of CC SNe are stripped of a fraction (SNe IIb) or all (SNe Ib) of their H and even He (SNe Ic) outer layers before explosion. 
The reason behind this sequence of different levels of mass-loss could be the initial zero-age main-sequence mass, metal content, rotation, or inter-layer mixing of the progenitor.
Two configurations may plausibly explain their nature. In the single-star scenario, the best candidates are massive (>25-30 \Msun) Wolf-Rayet stars that have been stripped of their envelopes by strong-line driven winds, which are dependent on metallicity \citep{2007ARA&A..45..177C,2001A&A...369..574V}.
The other possibility is lower-mass stars in binary systems that transfer the outer envelopes by interaction with the companion \citep{1992ApJ...391..246P}. In this latter case, the progenitor does not need to be as massive or as young as WR stars.
From the six stripped envelope SN progenitors detected in pre-explosion images, five are of SNe IIb \citep[e.g.][]{2004Natur.427..129M,2014ApJ...793L..22F}, all being consistent with binary progenitors, and only 1 SN Ib which was constrained to be a <25 \Msun~star also in a binary system \citep{2014AJ....148...68B,2015A&A...579A..95K}.
Therefore, questions such as the role of binarity, strong-line driven winds, and/or non-steady eruptive mass loss in determining SN type remain unanswered.

Probing the connection of observed supernova properties to a variety of possible progenitor stars has been a topic of interest in astrophysics for decades, but only for a few objects with pre-explosion Hubble Space Telescope imaging available has a direct connection been possible. 
Other methods include the analysis of SN spectra in the nebular phase (>200 days post-explosion) when the elements in the internal layers of the progenitor become optically thin to be detected \citep{2017arXiv170206702J,Kuncarakti15}.

The study of the SN environment, although not being a direct approach, has been able to find trends and put constraints by finding different degrees of association for SN types to a range of host environment properties (See \citealt{2015PASA...32...19A} for a review).
In this effort, integral field spectroscopy (IFS) has proved useful for both studies of individual nearby galaxies \citep{2017A&A...602A..85K} and of statistical samples of SN host galaxies in the nearby Universe.
Seminal works, such as \cite{S12}, showed the power of this approach and now IFS is the established approach for SN environmental studies 
\citep{2013AJ....146...30K,2013AJ....146...31K,2017arXiv171105765K,
2013A&A...560A..66R,
2014A&A...572A..38G,2016MNRAS.455.4087G,2016A&A...591A..48G,2017MNRAS.468..628G,
2018MNRAS.473.1359L}.

\begin{deluxetable*}{cccrl}
\tablecaption{Details of the observing programs in CAHA dedicated to obtain PMAS/PPak IFS data of SN host galaxies (PI: Galbany). \label{tab:prop}}
\tabletypesize{\scriptsize}
\tablenum{1}
\tablecolumns{4}
\tablewidth{0pt}
\tablehead{
\colhead{Semester} & \colhead{Proposal ID} & \colhead{Time awarded}  &\colhead{Objects observed}  & \colhead{Title} } 
\startdata
 15B & H15-3.5-004 & 4 nights  & 45 & IFS of core collapse supernova environments in low mass galaxies \\   
 16A & F16-3.5-006 & 5 nights  & 21 & Constraining supernova progenitors using the narrow Na absorption \\  
 16B & H16-3.5-012 & 2 nights  & 9 & Reducing Systematic effects in NIR SN Ia standardization \\  
 17A & F17-3.5-001 & 3 nights  & 12 & Reducing Systematic effects in NIR SN Ia standardization II\\ 
 17B & H17-3.5-001 & 2 nights  & 13 & Reducing Systematic effects in NIR SN Ia standardization III\\ 
\enddata
\end{deluxetable*}

This paper builds on previous work on SN environments within the Calar Alto Legacy Integral Field Area (CALIFA; \citealt{2016A&A...594A..36S}) survey and presents the PMAS-PPak Integral-field Supernova hosts COmpilation (PISCO) consisting of IFS data of 232 galaxies observed with PMAS/PPak and the 3.5m CAHA telescope using the same instrumental configuration. Observations for a 100 galaxies are presented here for the first time. 
PISCO is an effort that responds to the need of increasing the number of SN host galaxies observed with PMAS/PPak both (i) to construct statistical samples for SN subtypes not well represented in previous works, and (ii) to complete the CALIFA DR3 sample with objects in the low-mass regime, decreasing the host galaxy luminosity bias and probing a larger range in host stellar populations.
While specific studies regarding specific SN Ia-related topics will be published elsewhere, this paper aims to present the CC SN sample and update past results on correlation between SN type and star-formation and metallicity of its environment.
In addition, and taking advantage of the large amount of information contained in a single IFS galaxy observation, we performed an {\sc Hii} region statistical analysis which allows us to compare SNe locations with all other {\sc Hii} regions within their hosts.

In Section \ref{sec:sample} we describe the sample and give details on the specific programs that are part of PISCO.
Section \ref{sec:anal} contains a full explanation of the analysis performed on the datacubes that is used in following sections.
In Section \ref{sec:mass} we discuss the masses of the galaxies included in PISCO and how the new observations helped to reduce the previous existing bias towards high-mass galaxies present in the CALIFA survey.
In Section \ref{sec:local} we update previous results presented in \cite{2014A&A...572A..38G,2016A&A...591A..48G,2017MNRAS.468..628G} including all new observations from PISCO.
Section \ref{sec:hii} contains the results and discussion of a technique presented in \cite{2016MNRAS.455.4087G} that consists of performing {\sc Hii} region statistics and study the properties of the SN parent {\sc Hii} region in terms of all other {\sc Hii} regions in the galaxy.
In Section \ref{sec:prog} we discuss the implications of our results for different supernova type progenitors.
In Section \ref{sec:sniilocal} we present, for the first time, a correlation found with PISCO data between SN II post-maximum brightness decline and local star-formation rate intensity.
Finally, in Section \ref{sec:conc} we give a summary and list all our conclusions and in Section \ref{sec:future} outline future directions of work.


\begin{figure*}[p]
\centering
\includegraphics*[width=\hsize]{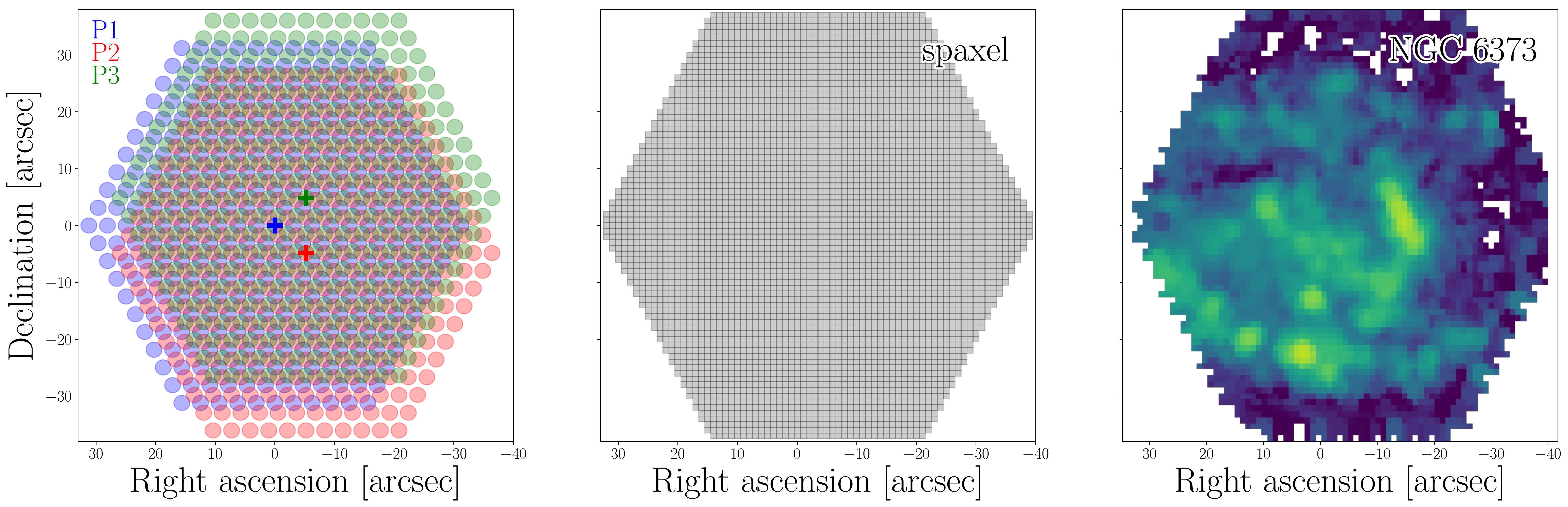}
\includegraphics*[trim=0cm 0cm 0cm -0.6cm, clip=true,width=\hsize]{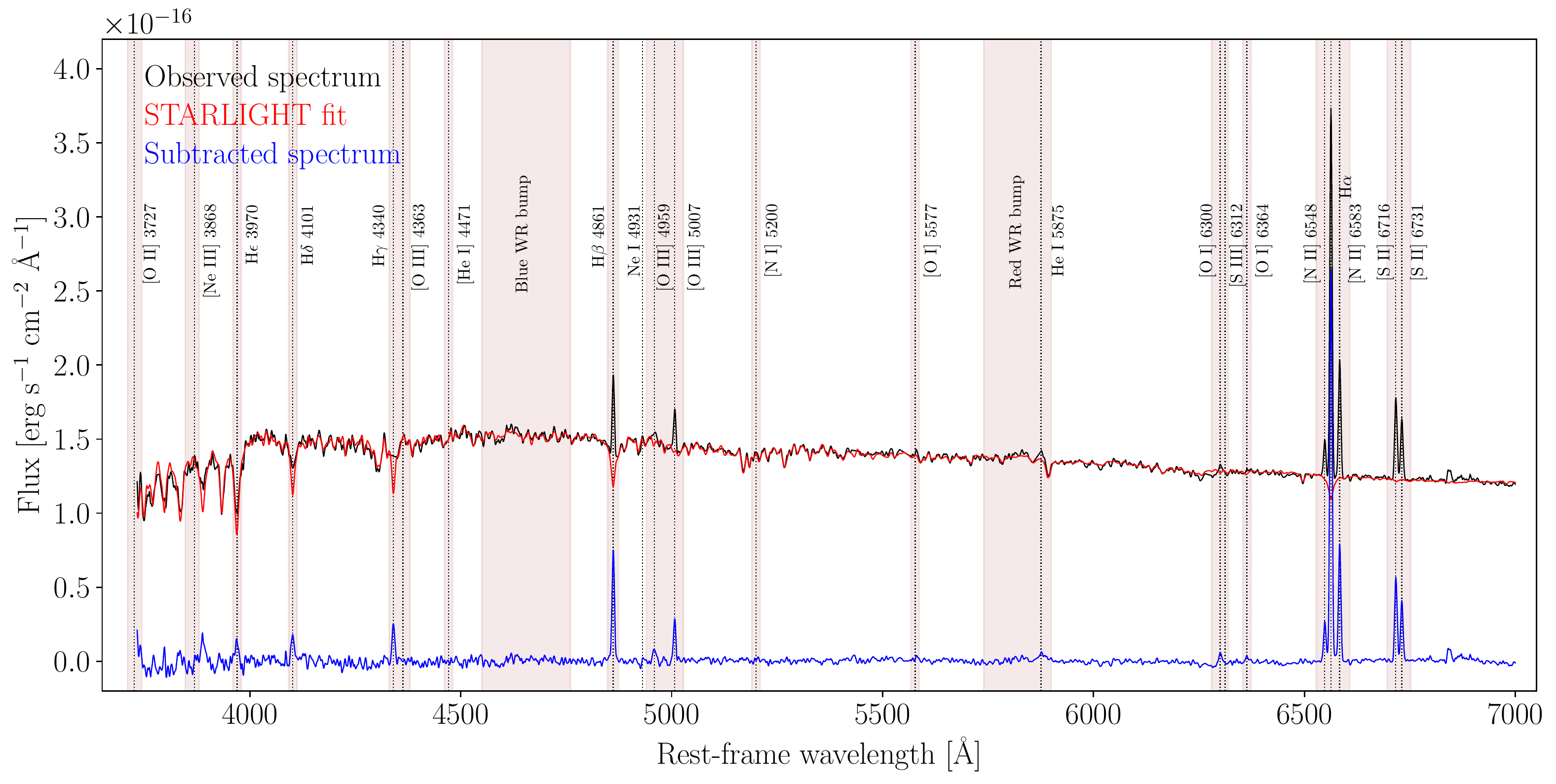}
\includegraphics*[trim=0cm 0cm 0cm 0cm, clip=true,width=\hsize]{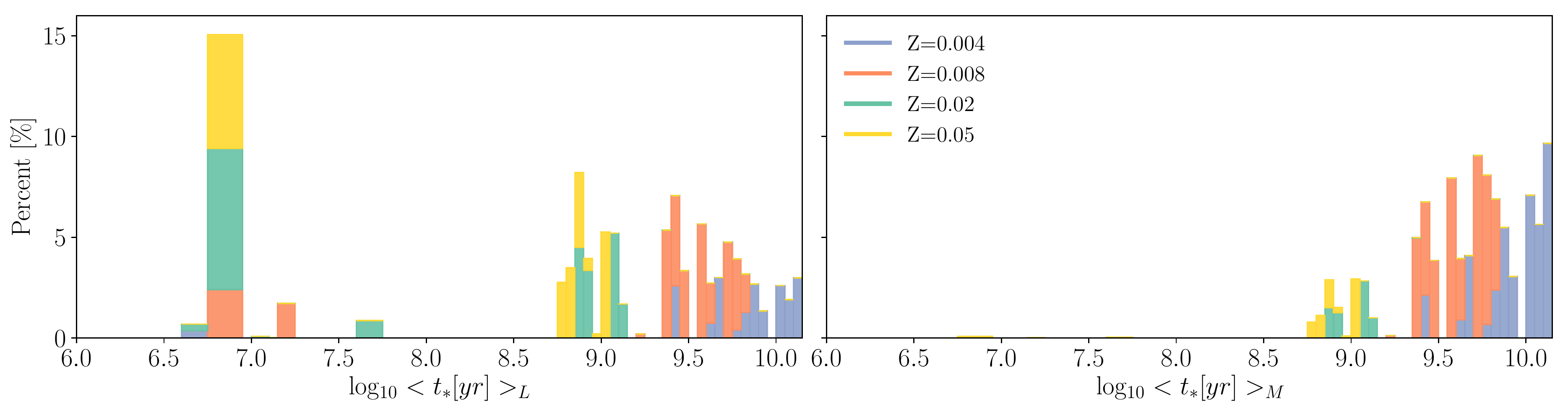}
\caption{{\it Top-left:} PPak science fiber-bundle showing the 3 pointings used to cover 100\% of the FoV. {\it Top-middle:} 1"-side spaxel configuration resulting of performing the dithering pattern. 
{\it Top-right:} Slice of a PMAS/PPak datacube corresponding to the position of the H$\alpha$ emission line.
{\it Middle:} Spectrum of NGC 6373 nucleus (black), together with the best STARLIGHT fit (red) and the pure nebular emission line spectrum
(the difference, in blue).  Brown shadows are the wavelength ranges masked in the stellar population fit. Vertical dotted lines correspond the positions of the fitted emission lines.
{\it Bottom:} Luminosity- and mass-weighted stellar age distribution of the fit shown in the middle panel. Contributions of the bases with different metallicities are displayed in different colors. }
\label{fig:ppak}
\end{figure*}

\vspace{0.5cm}
\section{Sample description} \label{sec:sample} 

The PMAS-PPak Integral-field Supernova hosts COmpilation (PISCO) puts together Integral Field Spectroscopy (IFS) observations of SN host galaxies performed with the Potsdam Multi Aperture Spectograph \citep[PMAS;][]{2005PASP..117..620R} in PPak mode  \citep{2004AN....325..151V, 2006PASP..118..129K} mounted to the 3.5m telescope of the Centro Astron\'omico Hispano Alem\'an (CAHA) located at Calar Alto Observatory in Almer\'ia, Spain.

PPak consists of a fiber bundle of 382 fibers each with a 2.7\arcsec~diameter, 331 of which (science fibers) are ordered in a single hexagonal bundle with a filling factor of the field-of-view (FoV) of 55\% (See Figure \ref{fig:ppak}; \citealt{2004AN....325..151V, 2006PASP..118..129K}). The remaining fibers are used for sky measurements (36), evenly distributed along a circle beyond the science fibers, and for calibration purposes (15).
Most of the observations were done with the V500 grating with 500 lines mm$^{-1}$ which provides a spectral resolution of $\sim$6~\AA\ and covers the whole optical range from 3750 to 7300~\AA.
A significant fraction (77/101) of the galaxies in PISCO were also observed with the higher-resolution V1200 grating, which has 1200 lines mm$^{-1}$ and covers the range from 3400 to 4750~\AA~with a resolution of $\sim$2.7~\AA. For those galaxies, we analyzed a combined cube from both the higher spectral resolution in the range covered by the V1200 grating, and the V500 resolution up to 7300~\AA.
Each galaxy was observed in three 900~s exposures. The second and third exposures were shifted by an offset of $\Delta(\mathrm{RA, Dec})=(-5.22, -4.84)$ and $(-5.22, +4.84)$~arcsec with respect to the position of the first exposure to ensure that every point within the FoV is spectroscopically sampled (See Figure \ref{fig:ppak} top-left panel). 
The combination of these three pointings provides wavelength- and flux-calibrated 3D datacubes with 100\% covering factor within a hexagonal FoV of $\sim$1.3 arcmin$^2$ with 1\arcsec$\times$1\arcsec pixels, which correspond to $\sim$4000 spectra per object.
For galaxies fainter than r$\sim$16 mag each individual exposure was increased to 1200s.

Around half of these observations (132 galaxies that hosted 154 SNe) were already presented in \cite{2014A&A...572A..38G,2016A&A...591A..48G,2017MNRAS.468..628G}, and were performed mostly in the framework of the Calar Alto Legacy Integral Field Area (CALIFA; \citealt{2012A&A...538A...8S,2014A&A...569A...1W}), although more specific details can be found in Appendix \ref{app:a}.
One of the {\it CALIFA-extension} projects described in the 3rd \cite{2016A&A...594A..36S} was {\it "IFS of core collapse supernova environments in low-mass galaxies"}, and contributed to the 3rd data release with 14 objects (marked in Table \ref{tab:sam} with $e$). Here we present the whole sample of SN host galaxies from that particular program, together with 4 new programs focused on SNe Ia, making a total of 100 SN host galaxies. 
The observing program code, title and time awarded for each of these 5 campaigns are summarized in Table \ref{tab:prop}. 

During the first campaign (semester 15B) we obtained observations of 45 (of the 50 proposed) low-mass ($<$10$^{10}$ \Msun) galaxies that hosted core-collapse supernovae (CCSN) with well-determined classification: SN types II, IIn, IIb, Ib, and Ic.
This project aimed to resolve a bias identified in \cite{2014A&A...572A..38G} due to the absence of low-mass galaxies in the CALIFA sample. The origin of this bias came from the construction of the sample, which was selected from the Sloan Digital Sky Survey (SDSS) DR7 with 2 simple cuts on redshift and angular size, hence discarding galaxies that did not cover a significant fraction of the instrument FoV \citep{2014A&A...569A...1W}.
With the addition of these new objects we are able to update the results presented in past works in terms of different SN subtypes and, in addition, increase the completeness of the SN host galaxy sample from the CALIFA Survey with  low-mass (< 10$^{10}$ \Msun) galaxies.
These SNe were selected from the Open Supernova Catalog (OSC\footnote{\href{https://sne.space}{https://sne.space}}; \citealt{2017ApJ...835...64G}) following these criteria: a) SN projected galactocentric distance lower than 40 arcsec, in order to cover the local SN environment with PMAS/PPak; b) Recession velocity of the galaxy lower than 9000 km/s ($\sim$z$<$0.03); c) $\log_{10}$ D25 (decimal logarithm of the apparent 25 mag/arcmin$^2$ isophotal diameter) lower than 1.12, which corresponds to galactic radius lower than 40 arcsec; d) Declination > 0 deg; e) SN light-curve publicly or privately available.

While specific studies regarding specific SN Ia-related topics will be published elsewhere, this paper aims to present the CC SN sample, update past results on correlation between SN type and star-formation and metallicity of its environment, and perform an {\sc Hii} region statistics analysis with the whole PISCO sample. 
Further details on the other programs can be found in Appendix \ref{app:b}. 

\begin{figure}[!t]
\centering
\includegraphics*[trim=0.15cm 0.3cm 0cm 0cm, clip=true,width=\hsize]{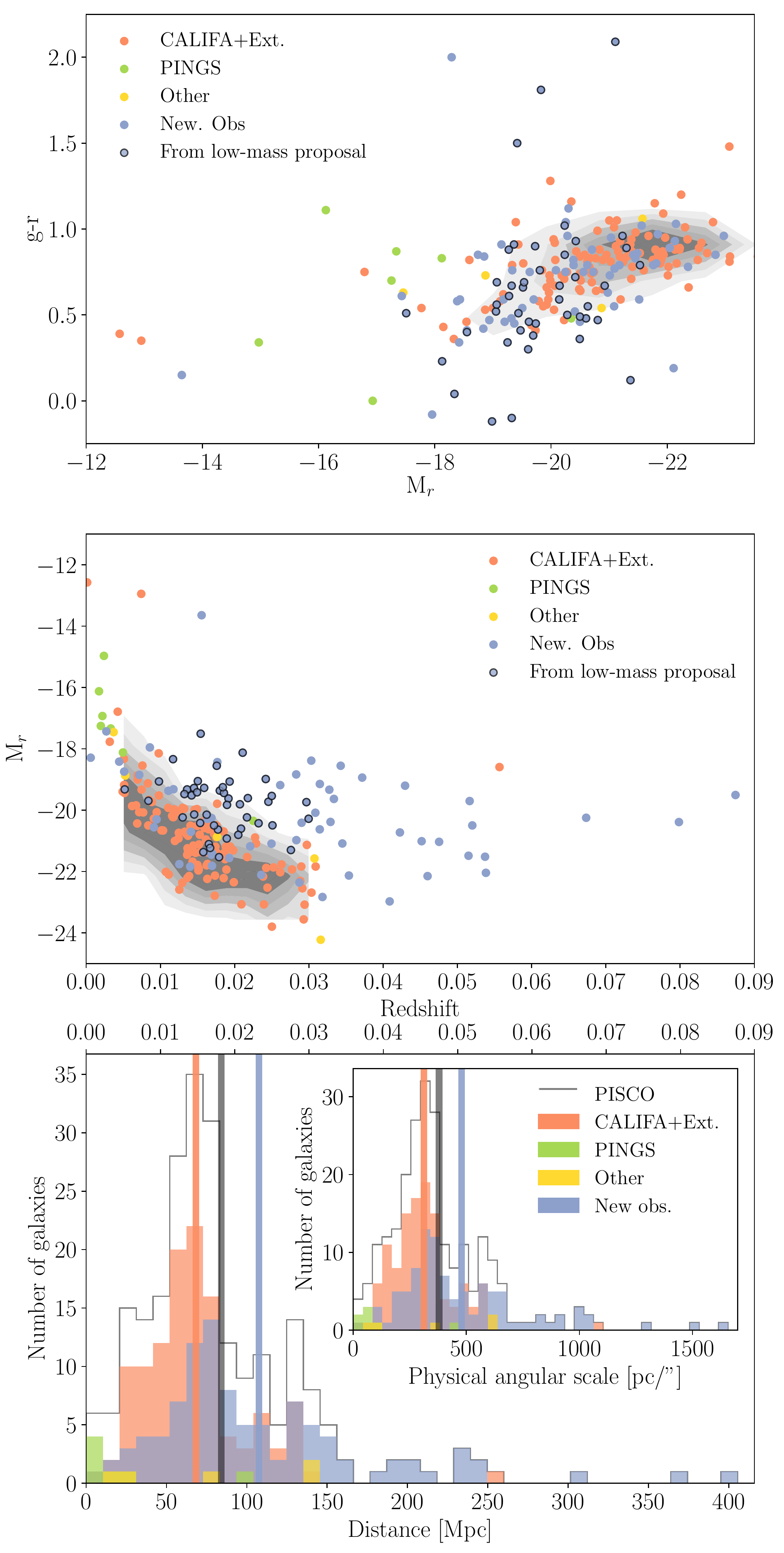}
\caption{{\it Top panel:} $g-r$ color $vs.$ absolute $r$-band magnitude ($M_r$) diagram.
On the background in grey contours are those parameters for the whole CALIFA sample as a reference. Blue dots with black contour correspond to those objects from PISCO that were observed under the low-mass galaxies program. 
{\it Middle panel:}
$M_r$ magnitude $vs.$ redshift for all galaxies included in PISCO colored by their source. Note that PINGS galaxies are more nearby than CALIFA DR3 galaxies, and galaxies presented here are fainter, therefore less massive, compared to the CALIFA  sample.  
{\it Bottom panel:} Distribution of redshift and physical angular scale (pc/spaxel; in the inner panel) for the 232 galaxies in PISCO. Vertical lines are the average redshifts for the CALIFA+Extensions sample (in red), the new observations presented here (in blue), and the whole PISCO sample (in black).}
\label{fig:dist}
\end{figure}

In Table \ref{tab:sam} we list the properties of the 232 galaxies and 272 SNe included in PISCO. This consists of 120 type Ia SNe (including two peculiar, two 91bg-like, six 91T-like, and one 02cx-like), 57 stripped-envelope SNe (including 19 Ib, 20 Ic, 12 IIb, one Ic-BL, four typed as Ibc, and one peculiar), and 95 type II SNe (including 19 IIn).
All galaxy parameters in Table \ref{tab:sam} are taken from the NASA/IPAC Extragalactic Database (NED\footnote{\href{https://ned.ipac.caltech.edu/}{https://ned.ipac.caltech.edu/}}) or the SIMBAD Astronomical Database\footnote{\href{http://simbad.u-strasbg.fr/simbad}{http://simbad.u-strasbg.fr/simbad}} when no information was available.
To reduce the effects of SN classification errors from initial discoveries because they might not be completely accurate, we performed a thorough search 
in discovery and classification reports from the Astronomers Telegram\footnote{\href{http://www.astronomerstelegram.org}{http://www.astronomerstelegram.org}} and the Transient Name Server (TNS\footnote{\href{https://wis-tns.weizmann.ac.il/}{https://wis-tns.weizmann.ac.il/}}), 
the Asiago \citep{1989A&AS...81..421B} and OSC SN catalogs, 
and in published literature through the NASA Astrophysics Data System (ADS\footnote{\href{http://www.adsabs.harvard.edu}{http://www.adsabs.harvard.edu}}) .
We tried to keep the most precise subtype classification (e.g. Ia 91bg-like, Ic-BL, etc...) when possible, but we left the more ambiguous classification when sources disagree or when we were not confident with the reclassification.

The two upper panels of Figure \ref{fig:dist} show a host galaxy $g-r$ color versus absolute $r$-band magnitudes and an absolute $r$-band magnitude $M_r$ versus redshift diagrams, all colored by the source the galaxy comes from. 
Galaxies from the CALIFA sample are included in grey contours for reference.
Galaxy magnitudes were retrieved from the SDSS DR14 ({\it modelmag}; \citealt{2017arXiv170709322A}), but for 46 objects we took their magnitudes from Pan-STARRS1 DR1 (\citealt{2016arXiv161205560C}; preferentially Kron magnitudes, but PSF when those were not available).
These panels clearly show how galaxies included in DR3 (labelled {\it CALIFA+Ext.} in Figure \ref{fig:dist}) lay on top of the CALIFA footprint, with some outliers that come from extension programs not following CALIFA sample cuts. Galaxies from PINGS are larger, brighter, and at  lower redshifts than the bulk of galaxies included in PISCO, as shown in the figure. 

We have plotted together (in blue) all 100 galaxies presented for the first time here, and it can be seen that those corresponding to the low-mass project (blue points with black edges) are objects at similar redshifts as the CALIFA sample but a few magnitudes fainter, which is explained by their smaller sizes (we did not require a minimum size in our sample selection). On the other hand, a few galaxies extend the redshift range up to 0.09, which correspond to SN Ia hosts observed in semesters 16B-17B (See Appendix \ref{app:b}).
In the lower panel of Figure \ref{fig:dist} we show the luminosity distance/redshift and angular physical scale (inner panel) distributions. 
The average redshift of all 232 galaxies included in PISCO is 0.0192, which corresponds to an angular scale of $\sim$380 pc/".


\section{Reduction and Analysis}\label{sec:anal} 

\subsection{Reduction}

All data was reduced with the same exact pipeline used for the CALIFA DR3 (v2.2; all details can be found in \citealt{2013A&A...549A..87H,2015A&A...576A.135G}, and \citealt{2016A&A...594A..36S}).
The reduction process is comprised of the following steps:
\begin{itemize}
\item[(i)] each set of four FITS files created by the amplifiers of the detector are re-arranged into a single frame, which is then bias subtracted and cleaned of cosmic rays;
\item[(ii)] relative offsets in the tracing due to flexure are estimated by comparing the continuum and arc-lamp calibration frames, the corresponding wavelength solution is applied to each individual 2D science frame, and all individual spectra are extracted using an optimal extraction algorithm \citep{1986PASP...98..609H} and stored in a row-stacked-spectrum (RSS) file; 
\item[(iii)] flux calibration is performed using a dedicated parallel program, which consists of reobserving two dozen CALIFA early-type galaxies (ETGs) and a set of the standard stars with the PMAS Lens-Array (LArr; Husemann et al. in prep.), and comparing the photometry from these observations to aperture-matched SDSS photometry in the g and r-bands;
\item[(iv)] science spectra corresponding to the three dithered exposures were combined into a single frame of 993 spectra. The flux corresponding to the 331 apertures of the fibers for each pointing is measured from sky-subtracted SDSS DR7 images in the bands covering the wavelength of our observation, when available. The apertures are shifted over a search box around the nominal coordinates of the pointing and the best registration is found on the basis of a $\chi^2$ comparison, which results in an accurate astrometry with a typical error of $\sim$0.2";
\item[(v)] a Galactic extinction correction is applied using dust maps from \cite{1998ApJ...500..525S} assuming a \cite{1989ApJ...345..245C} extinction law. {\it All cubes in PISCO include this correction in the reduced cubes};
\item[(vi)] a flux-conserving inverse-distance weighting scheme is used to reconstruct a spatial image with a sampling of 1" in a two-step process, first reconstructing the datacube and estimating the differential atmospheric refraction (DAR) offset, and secondly reconstructed the cube again but shifting the position of the fiber at each wavelength against the regular grid according to the DAR offset measured in the first step;
\item[(vii)] for the CALIFA DR3 data, when observations with two gratings are available a combined cube is created, which spans an unvignetted wavelength range of 3700$-$7140 \AA. The V1200 data is spatially recentered, flux rescaled and degraded in spectral resolution to match the V500 data, and the spectra corresponding to each dataset in the overlap region is averaged out, weighted by the inverse of the cube error;
\item[(viii)] finally, sky-subtracted SDSS DR7 g- and r-band images are downsampled to 1"/pixel (See Figure \ref{fig:ppak} top-middle panel), and corresponding g- and r-band images are also created from the CALIFA datacubes, to match their spectrophotometry as much as possible. Both images are registered using a Discrete Fourier Transform and the offsets are updated in the CALIFA image, and stored in the final 3D datacube. Following this procedure the absolute spectrophotometric accuracy is better than 3\%.
\end{itemize}

\begin{figure*}[p]
\centering
\includegraphics[trim=0cm 0cm 0cm 0cm,clip=true,width=\textwidth]{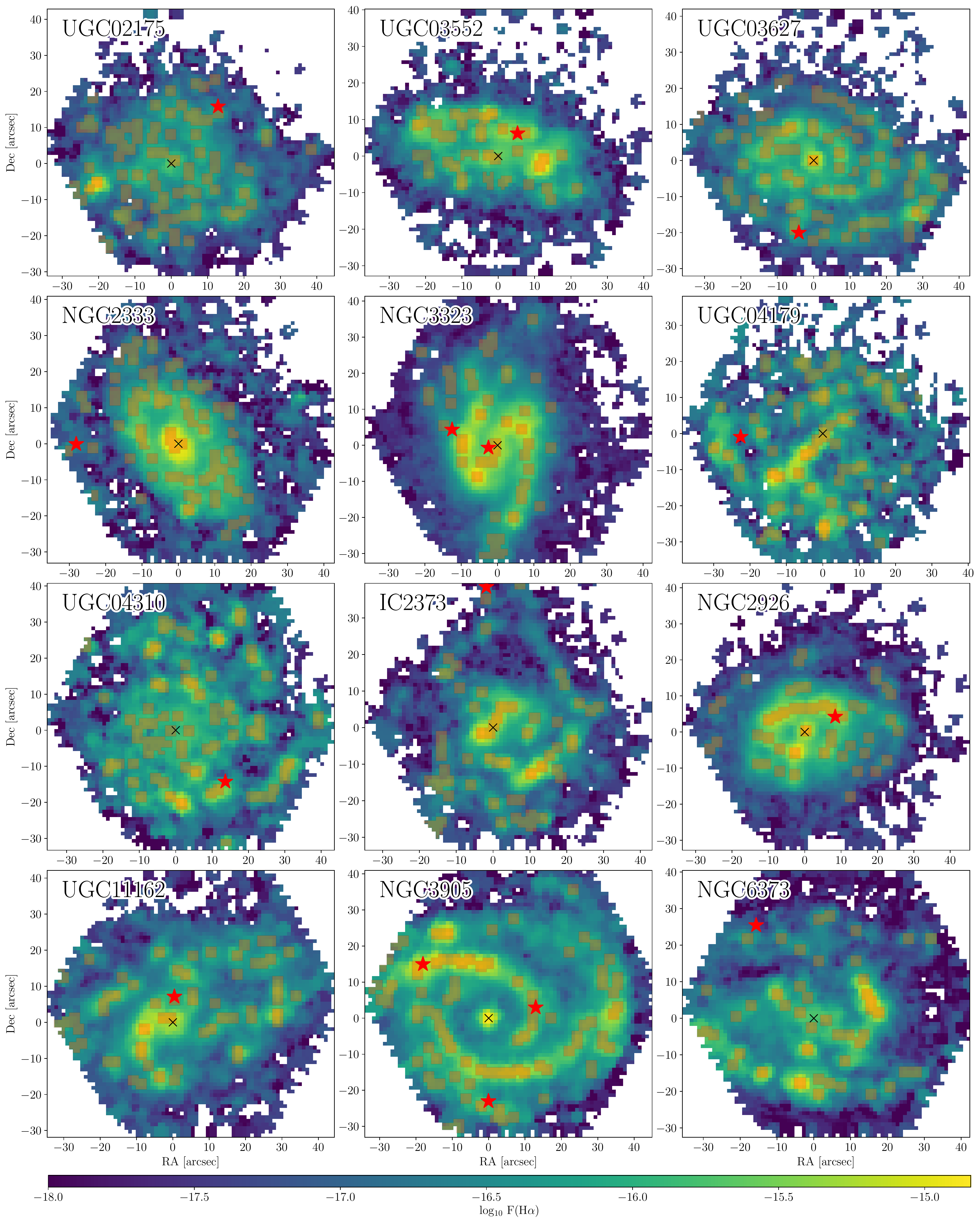}
\caption{Continuum-corrected H$\alpha$ emission line maps of 12 sample galaxies form the new observations listed in Table \ref{tab:sam}. Orange patches represent {\sc Hii} regions selected through the method described in Section \ref{sec:hiiexp}, red stars represent SN locations, and black $\times$ galaxy cores, which were measured by fitting a 2D-gaussian function to the collapsed map.}
\label{fig:maps}
\end{figure*}

\subsection{Analysis}

The analysis was performed in a similar way as that presented in \citealt{S12,2014A&A...572A..38G,2016A&A...591A..48G} and  \cite{2017MNRAS.468..628G}, employing our routines written in IDL\footnote{\href{http://www.harrisgeospatial.com/SoftwareTechnology/IDL.aspx}{http://www.harrisgeospatial.com/SoftwareTechnology/IDL.aspx}} and Python\footnote{\href{https://www.python.org/}{https://www.python.org/}}.
The procedures described below were applied to all individual 466,347 spectra included in the 232 datacubes, and also to the 71,654 spectra resulting from a Voronoi-binning tesselation \citep{2003MNRAS.342..345C} that substituted all spectra with S/N lower than 20 in a continuum band around 4800 \AA.
In addition, for each galaxy we extracted a spectrum in a circular aperture \mbox{1 kpc} in diameter centered on the galaxy core, and an integrated spectrum by summing up all spaxels in the FoV containing galaxy light with signal-to-noise ratio higher than 1. 
In order to exclude light from foreground stars we constructed 2D masks of point sources from SDSS or Pan-STARRS1 imaging.
The 2D spatial-configuration of our data allows us to characterize the nature of the emitted light from galaxies. As described below, we mapped regions whose emission is not dominated by ionization from star-forming regions but from the presence of Active Galactic Nuclei (AGN). We produced two different integrated spectra for those galaxies containing AGNs, the second excluding the above-mentioned region.
For SN locations, a circular aperture corresponding to 1 kpc diameter was also extracted. In case of low S/N, series of spectra were extracted in apertures centered at the SN positions and with radii up to 6\arcsec, and the spectrum with smallest aperture with S/N$>$3 in the same continuum band was used.

\subsubsection{Stellar populations}
We used {\sc STARLIGHT} \citep{2005MNRAS.358..363C,2009RMxAC..35..127C} to estimate the fractional contribution of different simple stellar populations (SSP) with different ages and metallicities to the stellar continuum in the spectra, assuming that the star formation history of a galaxy can be approximated as the sum of discrete star formation bursts (See Figure \ref{fig:ppak} middle and bottom panels). 
Dust effects, parametrized by A$_V^{\rm star}$ are modeled as a foreground screen with a \cite{1999PASP..111...63F} reddening law assuming R$_V$ = 3.1.
From the fits to the integrated spectra, the total stellar mass (M$_*$) is recovered by combining the mass-to-light ratio of the different SSPs contributing to the best fit. 
In this work we selected two different model bases:
\begin{itemize}
\item[(i)] to fit all individual and Voronoi spectra, we chose a selection of 66 components with 17 different ages (from 1~Myr to 18~Gyr) and four metallicities (0.2, 0.4, 1.0 and 2.5 \Zsun, where \Zsun=0.02) coming from a slightly modified version of the models of \cite{2003MNRAS.344.1000B}\footnote{See \cite{2007ASPC..374..303B} for more information.}, replacing STELIB by the MILES spectral library \citep{2006MNRAS.371..703S}, Padova 1994 evolutionary tracks, \cite{2003PASP..115..763C} initial mass function (IMF) truncated at 0.1 and 100~\Msun, with calculations of the TP-AGB evolutionary phase for stars of different mass and metallicity by \cite{2007A&A...469..239M} and \cite{2008A&A...482..883M}. 
\item[(ii)] to fit all integrated, central, SN location, and {\sc Hii} region (see Section \ref{sec:hiiexp}) spectra we used the ``Granada-Miles'' (GM) base, which is a combination of the MILES SSP spectra provided by \citet[][as updated by \citealp{2011A&A...532A..95F}]{2010MNRAS.404.1639V} for populations older than t = 63 Myr and the \cite{2005MNRAS.357..945G} models for younger ages. They are based on the \cite{1955ApJ...121..161S} Initial Mass Function and the evolutionary tracks by \cite{2000A&AS..141..371G}, except for the youngest ages (<3 Myr), which are based on Geneva tracks \cite{1992A&AS...96..269S,1993A&AS...98..523S,1993A&AS..102..339S,1993A&AS..101..415C}. The GM base is defined as a regular (t, Z) grid of 248 models with 62 ages spanning t = 0.001-14 Gyr and four metallicites (Z/Zsun = 0.2, 0.4, 1, and 1.5).
\end{itemize}
The main reason of choosing (i) over (ii) to do the initial analysis was the larger amount of spectra to fit ($\sim$500,000 over a few thousands) and the lower number of bases (66 over 248), which reduced the amount of computing time. 
In general, results from both bases are broadly similar as discussed in \citet{2015A&A...581A.103G}, although star formation histories are smoother in a larger base.

\subsubsection{Gas-phase emission}

We subtracted {\sc STARLIGHT} fits from the observed spectra to obtain pure gas emission spectra (See Figure \ref{fig:ppak} middle panel). We then accurately measure the flux of the most prominent emission lines by means of weighted nonlinear least-squares fit with a single Gaussian plus a linear term.
Errors on the flux measurement were determined from the S/N of the line flux and the ratio among the fitted amplitude to the standard deviation of the underlying continuum.
The flux of H$\alpha$ is a good proxy of the current ($<$10 Myr) star formation rate and has been extensively used in the literature  \citep{1998ARA&A..36..189K,2015A&A...584A..87C}.
Moreover, the intensity ratio of the observed H$\alpha$ (6563 \AA) and H$\beta$ (4861 \AA) emission lines, i.e., the Balmer decrement, provides an estimate of the optical extinction due to the dust attenuation. 
In photoionized nebulae, the intrinsic value of the I(H$\alpha$)/I(H$\beta$) ratio is 2.86, representative of low density nebular conditions $\sim$10$^3$ cm$^{-3}$ around a heating source with a typical temperature $T \sim 10^4$~K and large optical depths (Case B recombination; \citealt{2006agna.book.....O}).
Using a \cite{1999PASP..111...63F} extinction law, we obtain an estimate of the color excess \EBV~due to host galaxy dust which, adopting the Galactic average value of \mbox{\Rv = $A_V^{\rm gas}$ / \EBV = 3.1}, can be used to infer $A_V^{\rm gas}$. 
All optical emission lines previously measured are then corrected for the host galaxy dust extinction. 
In addition, H$\alpha$ equivalent widths (EW) are measured by first dividing the observed spectra by the STARLIGHT fit, and repeating the weighted nonlinear least-squares fit in the normalized spectra. 
Finally, 2D maps of emission-lines and stellar parameters were constructed.
Figure \ref{fig:maps} shows the extinction-corrected H$\alpha$ emission 2D maps for 12 of the new galaxies listed in Table \ref{tab:sam}. 
Fluxes and parameters measured in the local (1 kpc) SN spectra used in this work are reported in Table \ref{tab:measurements}.


\subsection{{\sc Hii} region segmentation}\label{sec:hiiexp}

Wide-field IFS allows SN explosion site parameters to be compared not only to those of the overall host, but to all other stellar populations found within hosts, therefore exploiting the full capabilities of the data.
Following \cite{2016MNRAS.455.4087G} we developed a method to characterize the SN parent cluster in comparison to all other {\sc Hii} regions in the galaxy, which consists of selecting the closest {\sc Hii} clump to the SN (under the assumption that the progenitor was born there) and then comparing such properties to the same parameters measured in all other nebular clusters in the galaxy.

Using our extinction-corrected H$\alpha$ maps we selected star-forming {\sc Hii} regions across each galaxy with {\sc HIIexplorer}\footnote{http://www.caha.es/sanchez/HII\_explorer/} \citep{2012A&A...546A...2S}, a package that detects clumps of higher intensity in a map by aggregating adjacent pixels until one of the following criteria is reached: 
(i) a minimum flux threshold (the median H$\alpha$ emission in the map), 
(ii) a relative flux with respect to the peak (10\% of the peak flux), 
or (iii) a radial distance limit (500pc).
The distance limit takes into account the typical size of {\sc Hii} regions of a few hundreds of parsecs (e.g., \citealt{1997ApJS..108..199G, 2011ApJ...731...91L}). 
The code starts with the brightest pixel, and iterates until no peak with a flux exceeding the median H$\alpha$ emission flux of the galaxy is left. 
Known caveats of the method are the following: the code does not always select individual {\sc Hii} regions since the physical scale of a real {\sc Hii} region could be significantly smaller than the CALIFA pixel size (1 to 6 regions per clump can be expected according to \citealt{2014A&A...561A.129M} );
and the code tends to select regions with similar sizes, although real {\sc Hii} regions have different sizes.

Once the {\sc Hii} regions were identified, the same analysis described above was performed to the extracted spectra. 
Spectra with EW(H$\alpha$) lower than 6~\AA~(justified by a correlation between EW(H$\alpha$) and the fraction of young populations; \citealt{2013A&A...554A..58S}, as shown in the WHAN diagram; \citealt{2011MNRAS.413.1687C}) or falling in the AGN region in the BPT diagram \citep{1981PASP...93....5B} according to \citep{2001ApJ...556..121K} criterium were discarded to make sure that the emission is caused by ionization from star-formation. 
From all 232 galaxies, the total number of {\sc Hii} regions that we are left with is 11,270, which makes an average of $\sim$49 per galaxy.
In Figure \ref{fig:maps} we show the resulting {\sc Hii} regions overploted on the H$\alpha$ 2D maps for 12 galaxies. 


\section{Galaxy stellar Masses}\label{sec:mass} 

\begin{figure}[!t]
\centering
\includegraphics[trim=0.0cm 0cm 0.0cm 0cm,clip=true,width=\columnwidth]{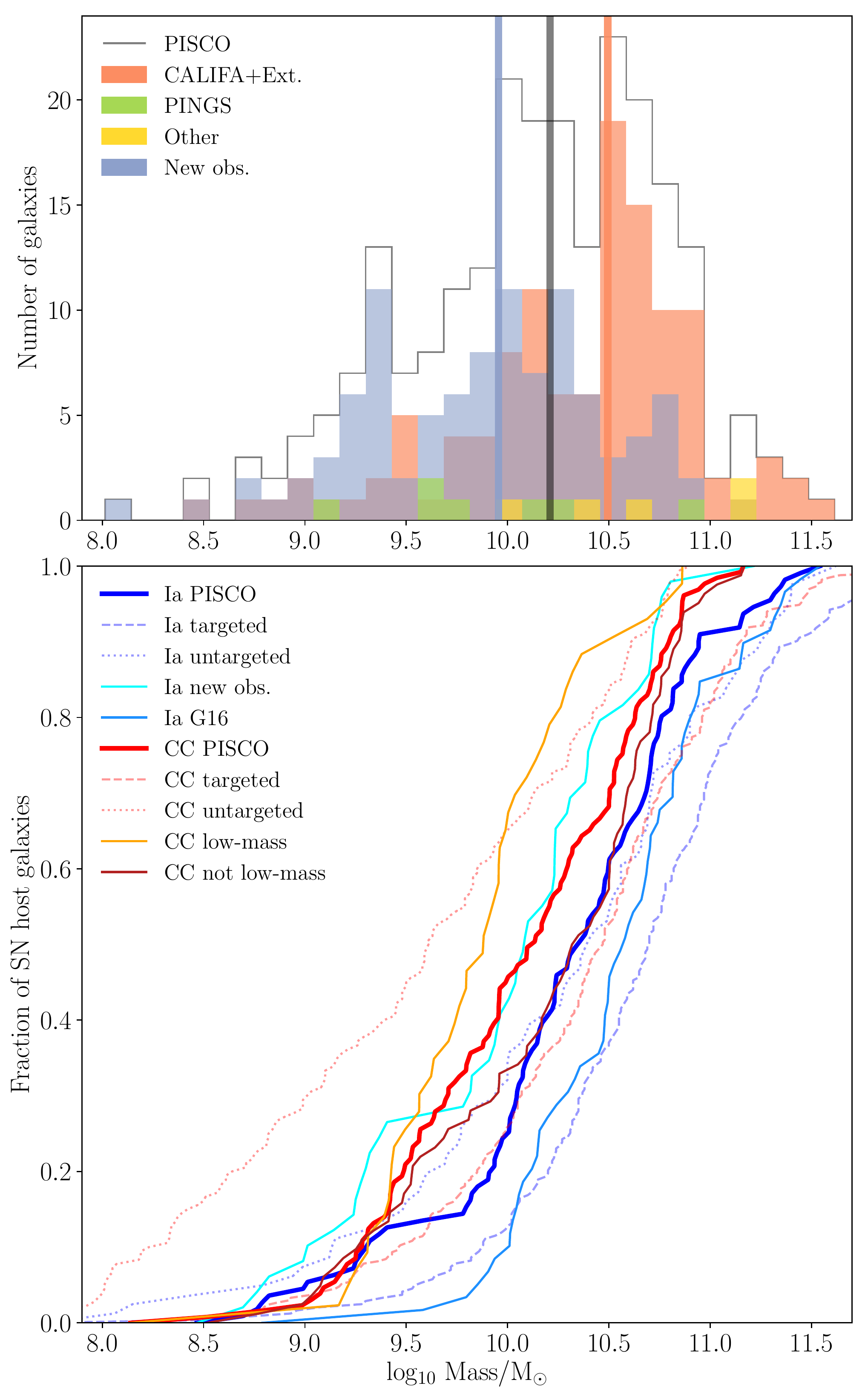}
\caption{Cumulative distribution of the masses of all PISCO galaxies separated by SN Ia and CC SN hosts. 
As a comparison we also show the untargeted (dotted lines) CC SNe hosts from the PTF \citep{2013ApJ...773...12S} and from \cite{2012ApJ...759..107K} compiled from several different surveys, as well as the targeted (dashed lines) CC SN galaxy samples of \cite{2012ApJ...759..107K}. 
We show untargeted SN~Ia hosts from PTF \citep{2014MNRAS.438.1391P} and targeted SN~Ia galaxies of \cite{2012ApJ...759..107K} and \cite{2009ApJ...707.1449N}.
This figure clearly shows that the SN targeted surveys are biased toward massive galaxies.
While in G16 we showed that the SN host galaxies from CALIFA had masses similar to other targeted searches, SN Ia host galaxies in PISCO follow the untargeted SN Ia host galaxy distribution. 
However, even with the addition of the galaxies observed under the low-mass project (orange distribution), we see that CC SN host galaxy distribution in PISCO is between CC SN targeted and untargeted distributions.}
\label{fig:mass}
\end{figure}

Figure \ref{fig:mass} shows the distribution of stellar masses, in units of log$_{10}$ \Msun, for various PISCO galaxy subsamples. In solid thick lines we show the distributions of all SN~Ia (in blue) and CC~SN (in red) host galaxies in PISCO. All mass measurements are reported in Table \ref{tab:measurements}. As expected SN Ia hosts have on average larger stellar masses (10.36 dex), as they include the earliest types, compared to CC SN hosts (10.11 dex) which are all late-type and also include a selected group of low-mass galaxies.

We also include, as a comparison in dotted and dashed lines respectively, stellar masses from representative targeted and untargeted SN host galaxy samples from the literature: 
targeted CC SN hosts of \cite{2012ApJ...759..107K};
untargeted CC SN hosts from the PTF survey \citep{2013ApJ...773...12S} and from \cite{2012ApJ...759..107K} compiled from several different surveys; 
untargeted SN~Ia hosts from PTF \citep{2014MNRAS.438.1391P};
and targeted SN~Ia galaxies of \cite{2012ApJ...759..107K} and \cite{2009ApJ...707.1449N}.
Finally, we split each PISCO SN Ia and CC SN subsample in two, to study how the representativeness of PISCO has improved compared with the sample presented in G16.
We performed two-sample Anderson-Darlington (AD) tests to check whether distributions were drawn from the same underlying population.
AD test is a modification of the Kolmogorov-Smirnov (KS) test that is more sensitive to the tails of distribution, whereas KS test is more sensible to the center of distribution.

SN Ia host galaxies presented in G16 have an average stellar mass of 10.58 dex and are representative of galaxies that host SNe Ia discovered by targeted searches (AD$_{\rm G16-Ia,T}$=0.10, AD$_{\rm G16-Ia,U}$=0.02).
In this present paper we add 38 new objects with an average lower stellar masses (10.08 dex). Figure \ref{fig:mass} clearly shows that these new objects convert the PISCO SN Ia galaxy sample into one that is compatible with the untargeted SN Ia PTF group (AD$_{\rm PISCO-Ia,U}$=0.29) and far from the targeted distribution (AD$_{\rm PISCO-Ia,T}$=1e-5).

For CC SNe, we added 44 objects from the low-mass project with significantly lower stellar masses (9.87 dex) on average than all the other CC SN hosts in PISCO (10.32 dex). These two distributions, in orange and dark red in Figure \ref{fig:mass}, are clearly shifted from each other (AD$_{\rm low,other}$=8e-4).
The G16 CC SN host galaxy sample was already not compatible with the targeted CC SN host compiled from the literature, but was more similar to that than to the untargeted sample (AD$_{\rm G16-CC,T}$=0.01, AD$_{\rm G16-CC,U}$=8e-6).
The addition of the new observations from the low-mass project shifts PISCO CC SN distribution to lower values, but still it is not compatible with the distribution of CC SN hosts discovered by untargeted surveys (AD$_{\rm PISCO-CC,U}$=1e-5), although it is now less compatible with the targeted distribution (AD$_{\rm PISCO-CC,T}$=2e-5).

The comparison of PISCO SN host properties with existing data for SN Ia and CC SN hosts leads us to conclude that we have been able to construct a SN Ia host galaxy sample compatible with SN Ia hosts from unbiased untargeted searches.
However, we are still far from a representative CC SN galaxy sample, although closer than in previous works.
This stresses the importance of continuing to obtain IFS data of galaxies with low masses --8.0 < log$_{10}$ (M [\Msun]) < 10.0-- in order to construct a sample of CC SN host galaxies free of biases introduced by targeted SN searches, which tend to discover objects in high-mass metal-rich hosts.

 \begin{deluxetable}{lcccc}
\tablenum{4}
\tablecolumns{9}
\tablewidth{0pt}
\tablecaption{Average values for each SN subtype \label{tab:averages}}
\tablehead{
\colhead{\textbf{SN type}} & \colhead{\textbf{log$_{10}$ ($\Sigma_{\rm SFR}$)}} & \colhead{\textbf{EW(H$\alpha$) [\AA]}} & \colhead{\textbf{$<$log$_{10}$ (t$_*$)$_L>$}} & \colhead{\textbf{12 + log$_{10}$ O/H}}} 
\startdata
\cline{1-5}    
Ia  & -2.317 $\pm$ 0.083 & 18.296 $\pm$  1.805 & 8.935 $\pm$ 0.060 & 8.643 $\pm$ 0.025 \\
II  & -1.809 $\pm$ 0.119 & 56.012 $\pm$  8.674 & 8.480 $\pm$ 0.070 & 8.540 $\pm$ 0.036 \\
Ibc & -1.795 $\pm$ 0.079 & 47.943 $\pm$  5.255 & 8.453 $\pm$ 0.071 & 8.577 $\pm$ 0.034 \\
Ib  & -1.635 $\pm$ 0.121 & 44.653 $\pm$  5.439 & 8.430 $\pm$ 0.103 & 8.623 $\pm$ 0.036 \\
Ic  & -1.671 $\pm$ 0.144 & 58.370 $\pm$ 12.168 & 8.386 $\pm$ 0.142 & 8.586 $\pm$ 0.070 \\
IIb & -2.264 $\pm$ 0.096 & 42.036 $\pm$  9.178 & 8.450 $\pm$ 0.140 & 8.388 $\pm$ 0.072 \\
IIn & -1.762 $\pm$ 0.141 & 48.389 $\pm$  6.284 & 8.401 $\pm$ 0.100 & 8.543 $\pm$ 0.055 \\
\enddata
\end{deluxetable}


\section{Local environmental parameters}\label{sec:local} 

\begin{figure*}
\centering
\includegraphics[trim=0cm 0cm 0cm 0cm,clip=true,width=\textwidth]{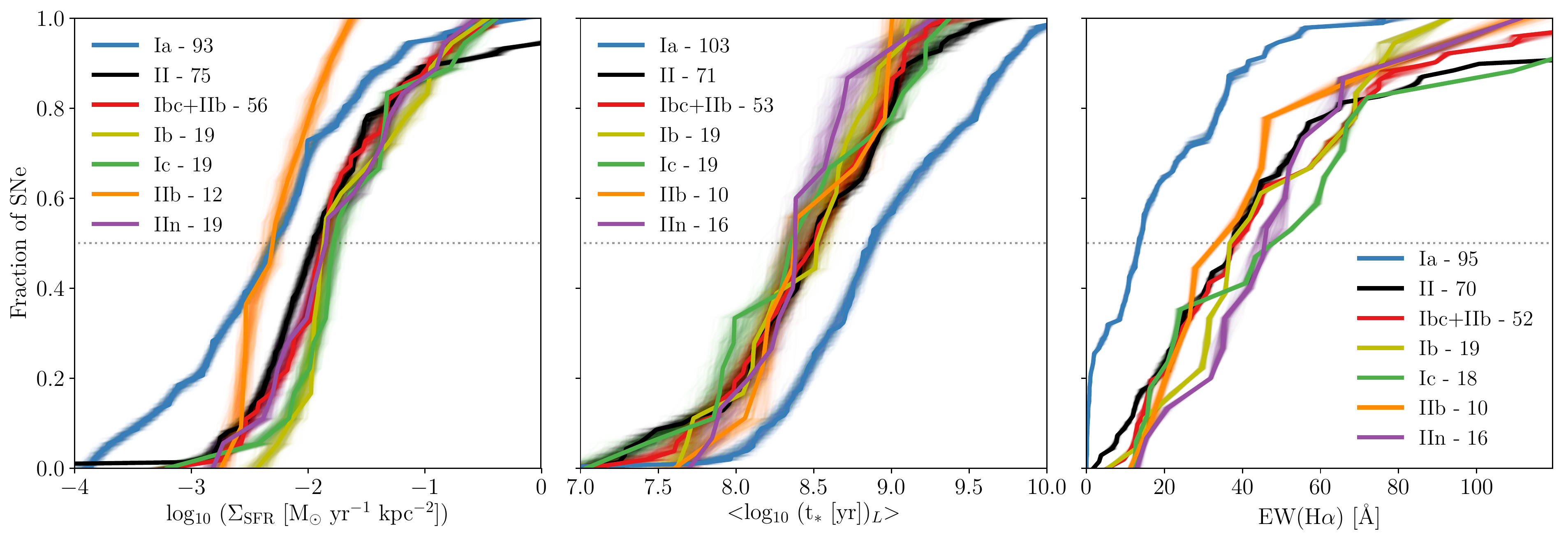}
\includegraphics[trim=0cm 0cm 0cm 0cm,clip=true,width=\textwidth]{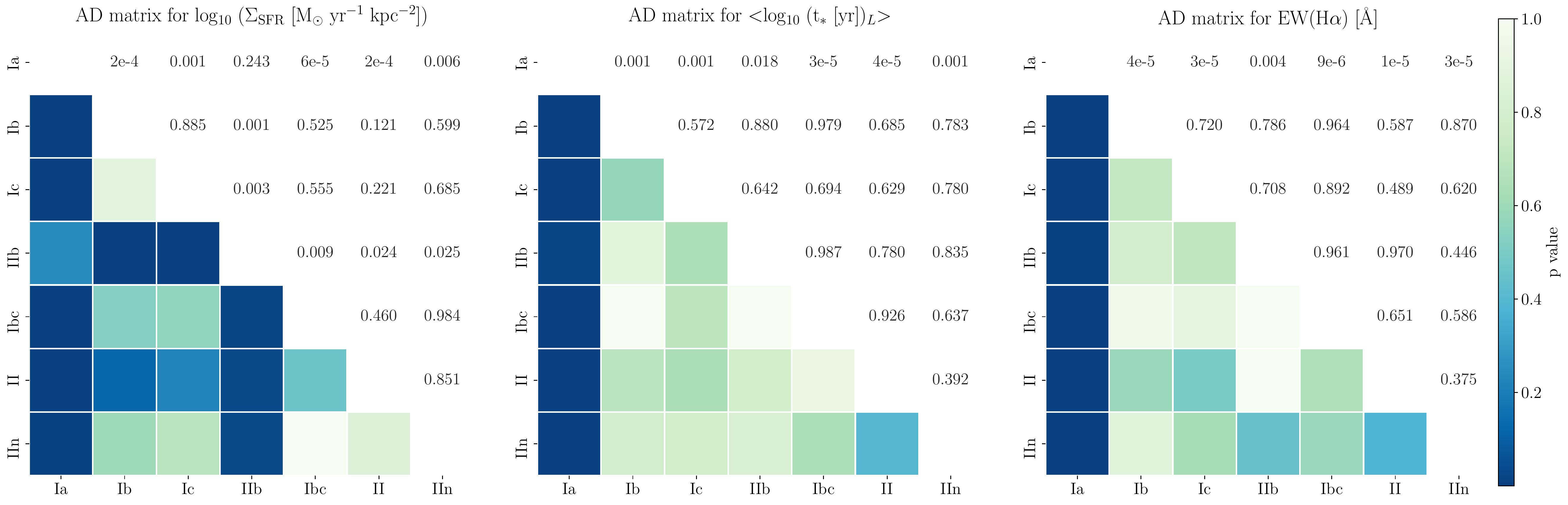}
\caption{Top row: Normalized cumulative distributions of the star formation rate intensity, luminosity-weighted average stellar age, and H$\alpha$ equivalent width measured in the local 1~kpc SN environment. A dotted horizontal line at a 0.5 fraction represent the median value of the distributions. Bottom row: Anderson-Darling statistic matrix of the three measurements for each combination of SN types.}
\label{fig:localparams}
\end{figure*}

In the following we study the differences between stellar and gas phase parameters measured in the spectra corresponding to the 1~kpc-diameter aperture centered at SN locations. 
We show the normalized cumulative distributions of star formation rate intensity ($\Sigma_{\rm SFR}$), H$\alpha$ equivalent width, the average stellar population age, and the oxygen abundance for the different SN types. In addition, we study how the star formation history (SFH) varies depending on the SN type found at those locations. These results update those presented in G14 and G16, which were obtained with a SN/galaxy sample less than half the size of the sample presented here. The increment in size has also allowed us to further split SNe in subtypes: from the 3 groups presented in G14 and G16 (Ia, II, and Ibc), we are now able to differentiate between SN II with signs of interaction (SN IIn) and other `normal' SNe II (including historical types L and P), and between types Ib, Ic and IIb independently.
However, for consistency with our previous works, we still include the {\it Ibc+IIb} group in our plots, which contain SNe that were accurately classified as Ic, Ib, or IIb, together with other stripped-envelope SNe with less accurate classification, such as Ibc or peculiar.

For visualization purposes of the individual measurement errors, when constructing normalized cumulative distributions shown in Figures \ref{fig:localparams}, \ref{fig:localoh} and \ref{fig:ohNT}, we generated 1000 distributions for each SN subtype by randomly sampling each measurement using a normal distribution centered at the measured value and with the error as one sigma.
In those Figures and in Figure \ref{fig:localpercents} we also summarize the results of the two-sample AD tests performed among all distributions.
Darker-blue colors represent lower p-values which correspond to distributions having underlying populations that are significantly different from each other, while brighter-greenish colors correspond to distributions that are similar to each other.  
Average values of all parameters for each SN subtype are listed in Table \ref{tab:averages}.

\subsection{Correlation with star formation}

The extinction-corrected H$\alpha$ flux $F(\mathrm{H}\alpha)$ is broadly used as a proxy of the the ongoing star formation rate (SFR; \citealt{1998ApJ...498..541K}).
\cite{2015A&A...584A..87C} demonstrated that the H$\alpha$ luminosity alone can be used as a tracer of the current SFR, even without including UV and IR measurements, if the underlying stellar absorption and the dust attenuation effects have been accounted for, which is the case of this work.
In order to properly compare results from galaxies at different redshifts, the fairest parameter is the SFR intensity ($\Sigma_{\rm SFR}$=SFR/area). Given that the spectra used in this section have the same physical aperture in all cases (1 kpc), this effect has already taken into account and our SFR measurements are already SFR intensities.

In the left panel in Figure \ref{fig:localparams} we show distributions of $\Sigma_{\rm SFR}$ for all SN subtypes. For consistency, we also plotted together the stripped envelope SNe distribution (Ib+Ic+IIb, in red).
The SN~Ia distribution shows the lowest average value. Although SN Ia occur at both high and low $\Sigma_{\rm SFR}$, it is the only type occurring at locations with $\log_{10}$ $\Sigma_{\rm SFR}$ < -3 dex.
Most of the CC SN types are clustered at higher values, but the SN IIb distribution is significantly shifted to lower SFR intensity.
Moreover, the steepness of SN IIb distribution compared to all other distributions also indicates that it is the narrowest, with no environments with $\log_{10}$ $\Sigma_{\rm SFR}$ $\gtrsim$ -2 dex.

\subsection{Correlation with stellar age}

The SSP fitting performed with STARLIGHT provides a reconstruction of the star formation history (SFH). 
We here calculate the luminosity-weighted average stellar age in all 1~kpc local spectra, and show the cumulative distributions in the middle panel of Figure \ref{fig:localparams} for each SN type.
Note that this measurement is built from information of the continuum and does not take into account the gas-phase emission lines at all.

The only significant difference between the distributions, as expected, is that SNe Ia tend to happen at locations where the average age of the stellar populations is older.
CC SN distributions are clustered at young ages, although the average values are ordered in a sequence from SN Ic, IIn, II, IIb and Ib in increasing average age.

\subsubsection{H$\alpha$ equivalent width}

While the H$\alpha$ line luminosity is an indicator of the ongoing SFR traced by ionizing OB stars, the H$\alpha$ equivalent width (EW) measures how strong this is compared to the continuum, which is dominated by old low-mass non ionizing stars and therefore accounts for most of the galaxy stellar mass. 
EW(H$\alpha$) can be thought of as an indicator of the strength of the ongoing SFR compared with the past SFR, which reduces with time if no new stars are created, and therefore it is a reliable proxy for the age of the youngest stellar components \citep{2016A&A...593A..78K}.

EW(H$\alpha$) can be calibrated to stellar age using SSP models (i.e., Starburst99; \citealt{1999ApJS..123....3L}) assuming single stars born in an instantaneous star formation and distributed in mass according to a particular initial mass function (IMF).
Under these assumptions, {\sc Hii} regions are no longer bright in H$\alpha$ after $\sim$10 Myr.
However, if either continuous star formation bursts are considered or  binary stellar populations included, 
the ionizing populations can survive further extending 
the lifetime of H$\alpha$ emission and therefore the corresponding stellar ages up to $\sim$100 Myr \citep{2009MNRAS.400.1019E,2011AJ....141..126G,2017A&A...601A..29Z}.

EW(H$\alpha$) is the parameter where the clearest differences are found between different CC SN distributions, as can be seen in the right panel of Figure \ref{fig:localparams}. The SN~Ic distribution has the highest values, followed by the SNe IIn, II and Ib distributions. Again, SNe IIb have significantly lower values than other CC SNe. 
More than 20\% of SNe Ia occur at locations with no or insignificant EW(H$\alpha$), which might mean that there is no gas emission at all, or that the continuum dominates the detected light by a large factor.

\subsubsection{Star formation histories}\label{sec:sfh}

\begin{figure*}
\centering
\includegraphics[trim=0cm 0cm 0cm 0cm,clip=true,width=\textwidth]{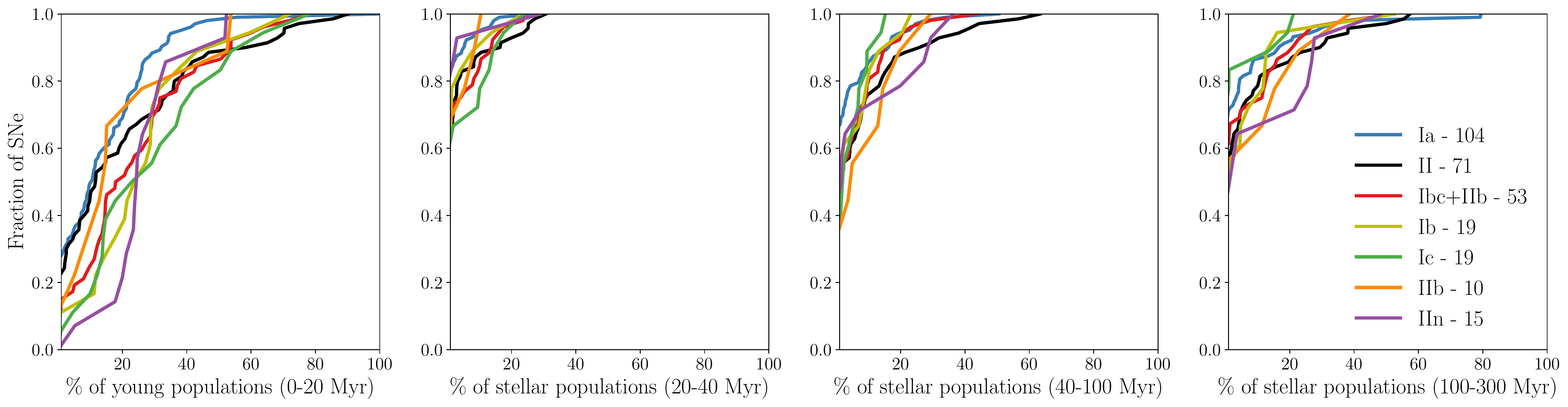}
\includegraphics[trim=0cm 0cm 0cm -1cm,clip=true,width=0.9\textwidth]{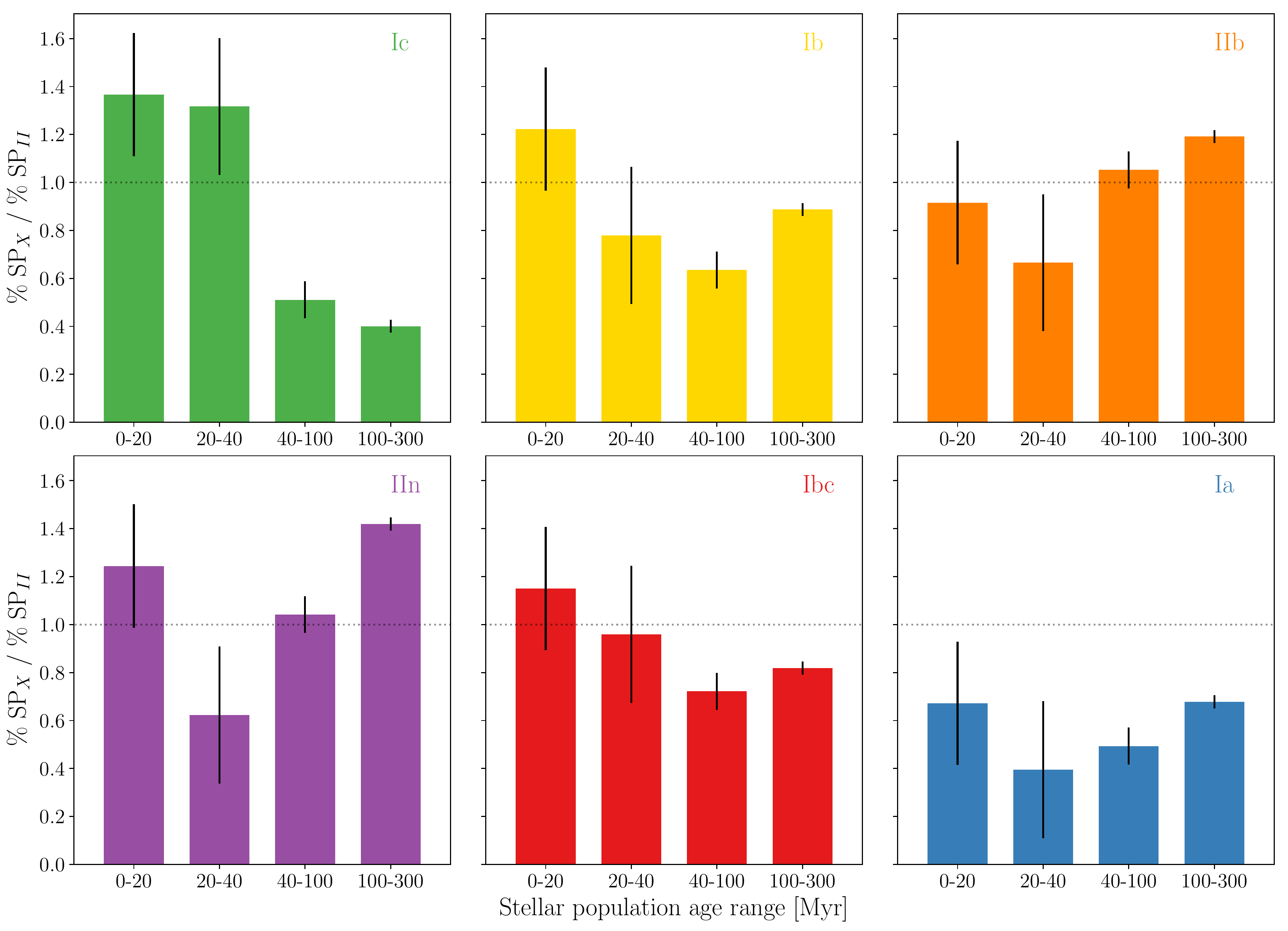}
\caption{ Top: Normalized cumulative distributions of the luminosity-weighted percentages of models used in the best STARLIGHT fit divided in 4 age bins of ages (0-20, 20-40, 40-100, 100-300 Myr). 
Bottom: Distributions of the average star formation histories divided in those 4 age bins relative to the average percentages of the SN II distribution (black distribution in the panels above and throughout the paper).}
\label{fig:localpercents}
\end{figure*}

Going one step further from the analysis of average stellar ages, and given that SSP fitting allows reconstruction of the star formation history, here we study the stellar populations younger than 300 Myr present at SN locations.
From the 248 models available in the ``Granada'' base used in the SSP fitting, 112 models correspond to populations of interest ($<$300 Myr). We divided these models in 4 age bins: 0 to 20 Myr (40 models), 20 to 40 Myrs (12 models), 40 to 100 Myr (24 models), and 100 to 300 Myr (36 models).
Note that Granada base have four different metallicities for each age, so the number of models with different ages in each age bin is four times smaller.

Figure \ref{fig:localpercents} shows the normalized cumulative distributions of the percentage of models in that particular age bin that were needed to reconstruct the observed spectra in STARLIGHT. The middle panel in Fig. \ref{fig:localpercents} shows the AD matrix.
The distributions of the youngest age bin (upper-left plot) show a similar sequence to the parameters studied at previous sections: SN Ic locations tend to need more populations in this age bin, followed by SNe Ib and SNe IIn, whose distribution is significantly different than SNe~II. Again, the SN IIb distribution differs from the other two stripped-envelope SNe, which make the common SNe Ibc-IIb distribution (red line) be more similar to SNe II, than the individual distributions. Finally, SNe Ia locations require a lower percentage of very young populations, as compared to CC SNe.
In the other three panels we note how the SN Ic distribution shifts to the left as the age bin goes to older ages, and SNe IIb and IIn are the subtypes needing additional components of older ages.

\begin{figure*}
\centering
\includegraphics[trim=0cm 0cm 0cm 0cm,clip=true,width=\textwidth]{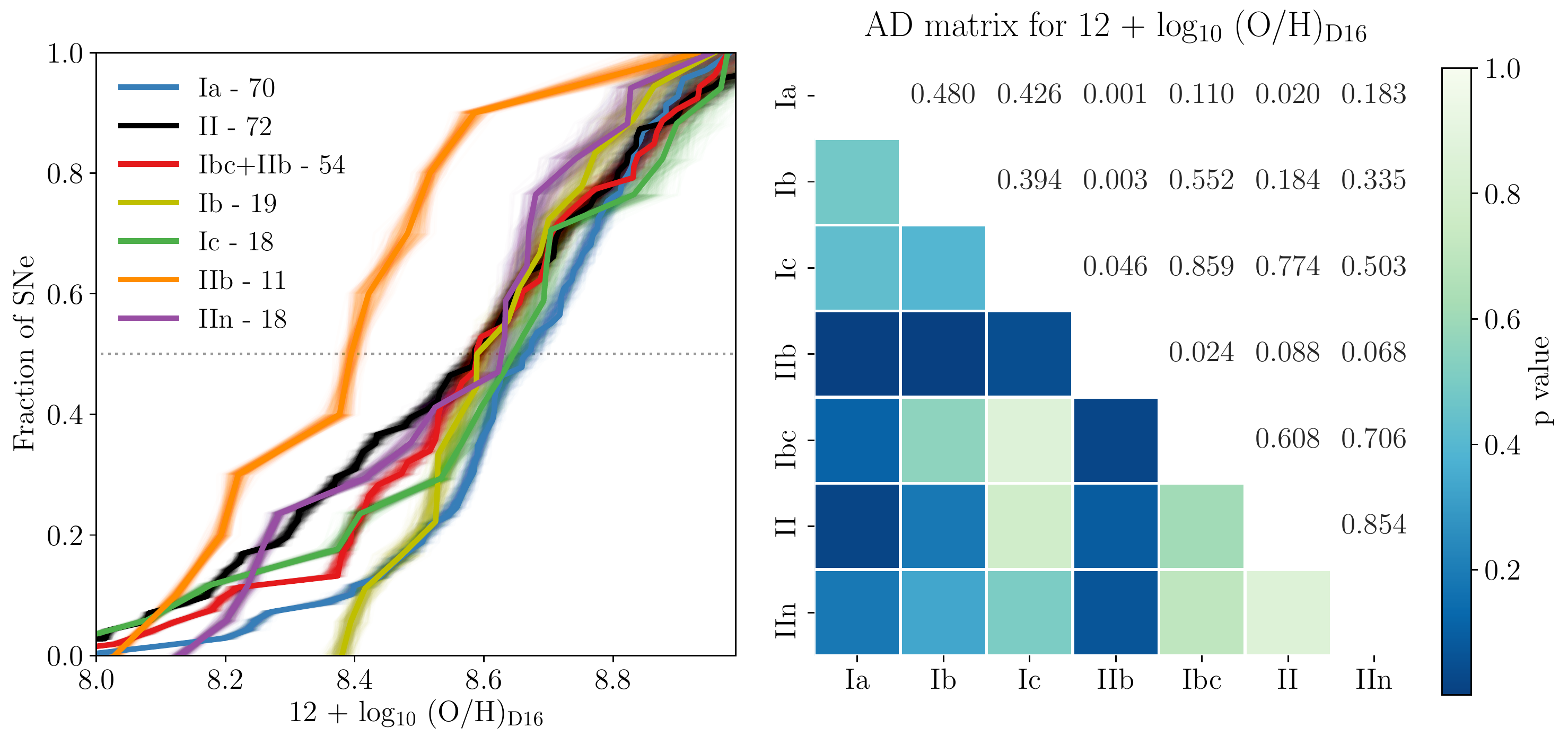}
\caption{Left: Normalized cumulative distributions of the oxygen abundance measured in the local 1 kpc SN environment. A dotted horizontal line at a 0.5 fraction represent the median value of the distributions. Right: Anderson-Darling statistic matrix for each combination of SN types.}
\label{fig:localoh}
\end{figure*}

Bottom panel of Figure \ref{fig:localpercents} puts together all the information shown in the upper panels, by normalizing the percentage in each bin to the percentage of  SNe II (black lines in the upper panels). In this way, given the evidence of RSG being the progenitors of SNe II, we can see how the percentages of other SN types evolve as the age of the stellar populations gets older.
All trends pointed out above are clearly seen in this Figure:
\begin{itemize}
\item SNe Ic (green) is above the normalization factor in the first two bins and goes below in the two older bins. 
This has to be read in terms of "SNe Ic have on average more young stellar populations of 0-40 Myr than SNe II".
\item On the other hand, the SN Ib (yellow) evolution is flatter in comparison to SNe Ic. While in the innermost bin SNe Ib require more young populations than SNe II, the average in others bins is below. In general terms SN Ib behaviour is quite similar to the Ibc/IIb group (in red). 
\item SNe IIb (orange) requires more contribution of populations older than 40 Myr and less younger than 40 Mys than SNe II. This is the opposite behaviour than SNe Ic.
\item Another interesting result is that SNe IIn (in purple) have a bimodal behavior, with higher averages both in the innermost and in the older bins and lower or similar percentages than SNe II in the central two bins. 
\end{itemize}

\subsection{Correlation with metallicity}

\begin{figure}
\centering
\includegraphics[trim=0cm 0cm 0cm 0cm,clip=true,width=\columnwidth]{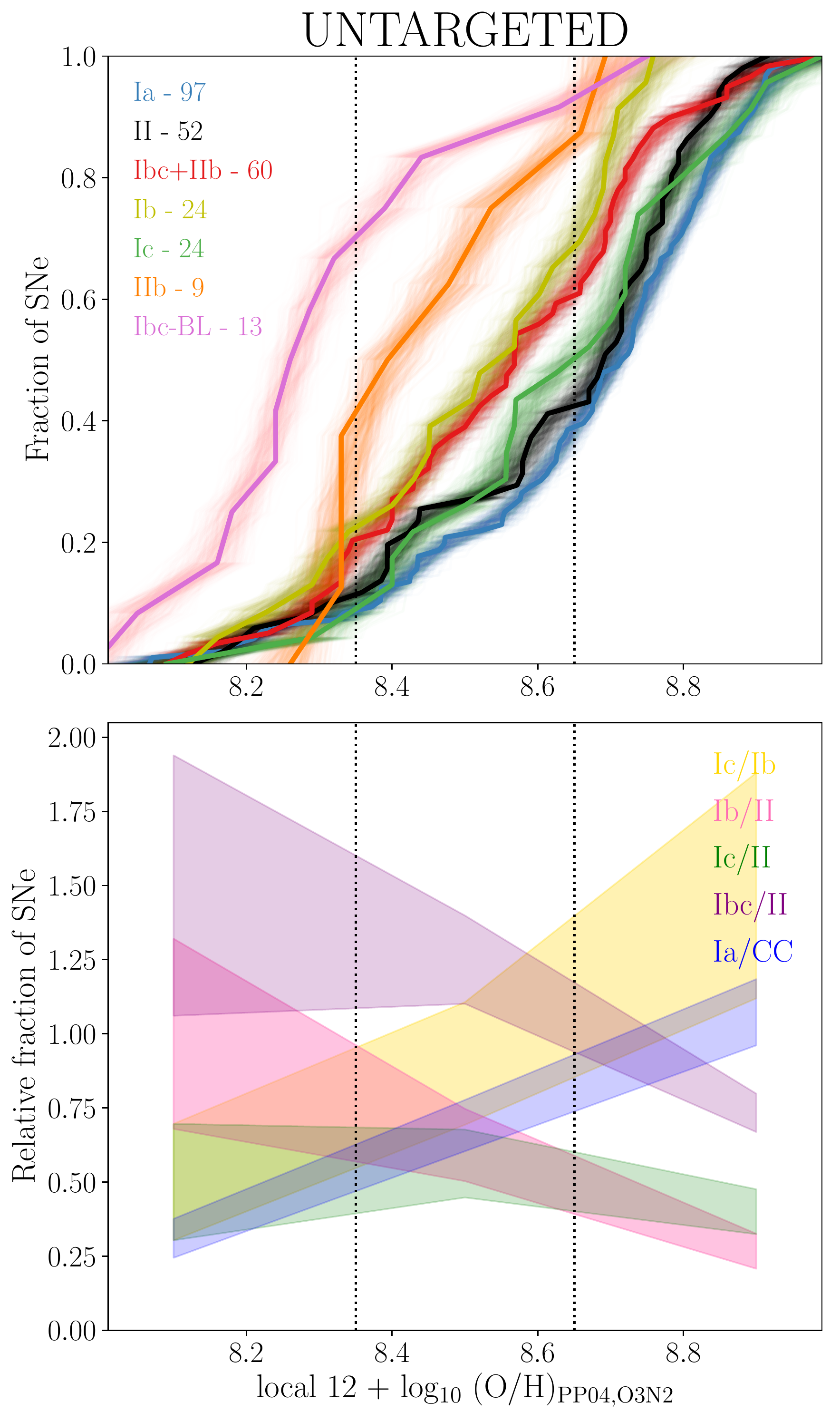}
\caption{Gas metal abundance measurements at SN locations from PISCO and the literature. Here we only include SNe that have been discovered by untargeted searches. On top, we show the normalized cumulative distributions for each SN subtype, and at the bottom panel we shows how several number ratios evolve as metallicity increase. We calculated the ratio in three bins (limits at 8.35 and 8.65): sub-solar, solar, and over-solar. Colored strips connect dispersions measured in each bin.}
\label{fig:ohNT}
\end{figure}

Oxygen is the most abundant metal in the gas phase and exhibits very strong nebular lines at optical wavelengths. This is why it is usually chosen as a metallicity indicator in interstellar medium (ISM) studies.
The most accurate method to measure ISM abundances (the so-called direct method) involves determining the electron temperature of ionized gas, T$_e$, which is usually 
directly measured from the temperature sensitive intensity ratios of collisionally excited forbidden lines (e.g., \mbox{[O III] $\lambda$4363/$\lambda$5007}; see \citealt{2006agna.book.....O}), followed by an analysis of the various ionization fractions in the zones of the {\sc HIii} region which produce optical emission lines.  This has been very extensively applied by observers over the past 30 years (e.g., \citealt{1969BOTT....5....3P,1978A&A....66..257S,1987MNRAS.226...19D,1996MNRAS.280..720V,2009A&A...508..615L}).
Temperature is anticorrelated with abundance which makes auroral lines disappear in metal-richer environments, so other strong-line emission methods have to be used instead.
While theoretical methods are calibrated by matching the observed line fluxes with those predicted by theoretical photoionization models, empirical methods are calibrated against H{\sc II} regions and galaxies whose metallicities have been previously determined by the direct method.
Unfortunately, there are large systematic differences between methods, which translate into a considerable uncertainty in the absolute metallicity scale (see \citealt{2012MNRAS.426.2630L} for a review), while relative metallicities generally agree.
The cause of these discrepancies is still not well-understood, although the empirical methods may underestimate the metallicity by a few tenths of dex, while the theoretical methods overestimate it \citep{2007RMxAC..29...72P, 2010ApJS..190..233M}. 

Figure \ref{fig:localoh} shows normalized cumulative distributions and AD matrix for the local oxygen abundance measured using the \citet[D16]{2016Ap&SS.361...61D} calibrator based on photoionization models, but similar relative results stand when the \cite{2013A&A...559A.114M} O3N2 empirical calibrator is used.
Both methods have the advantage of being insensitive to flux calibration and extinction due to the small separation in wavelength of the emission lines used for the ratio diagnostics, and neither suffer from differential atmospheric refraction (although this is already corrected for when reconstructing the IFS datacube; see Section \ref{sec:anal}).
While we are not going to strongly rely on the absolute numbers, in the following we will discuss the results in a relative sense. 

SNe Ia in PISCO tend to occur in metal-richer environments more so than all other core-collapse SN types, however the difference is significant only when compared to non-interacting hydrogen rich (type II) SNe and SNe IIb.
In fact, SN IIb occur at significantly metal-poorer locations compared to all other SN types except SNe II and IIn (although AD is 0.09 and 0.07, respectively).
The differences among SNe Ic, Ib, and II are insignificant. Indeed, we noted already in \cite{2016A&A...591A..48G} that differences among these types arise only when  SNe discovered by untargeted searches are used. Otherwise, the results are biased towards metal-rich galaxies which are those observed by targeted searches.
From the 272 SNe included in PISCO, 205 were discovered in targeted searches and 62 by untargeted surveys. 
When adding all new observations from PISCO to the same compilation of SNe from untargeted searches from the literature presented in \cite{2016A&A...591A..48G}\footnote{The sample studied here has the following differences with that used in \cite{2016A&A...591A..48G}: (i) we here do not include those SNe Ia measurements that were performed at the center of their host galaxies but then corrected by an offset of 0.029 dex to account for the difference between local and central values; (ii) we include the measurement of SN 2017ahn from \cite{2017arXiv171105765K}, which was a single object from an untargeted search; and (iii) we disregarded measurements of SNe environments from other works that are now part of PISCO.}, we recover the same sequence in environmental metallicity from SNe Ia to broad-line SNe Ic. 
In the upper panel of Figure \ref{fig:ohNT} all cumulative distributions are presented. Most of the measurements from the literature were reported in the O3N2 \cite{2004MNRAS.348L..59P} scale, so we converted other measured in a different scale to O3N2. Two dotted vertical lines represent the two breaking points used to divide in three bins, and the number ratios of SNe of different types in those three bins are presented in the lower panel of Figure \ref{fig:ohNT}.

Two important facts are shown in these ratios:
(i) Even when we have compiled a sample free of objects discovered from targeted searches, it is not volume-limited and other biases can be at play (e.g., our search of CC SNe in low-mass galaxies may be increasing the number of stripped-envelope SNe at low metallicities, as shown by the Ibc/II ratio in Fig. \ref{fig:ohNT} bottom panel), so differences between SNe II and stripped envelope SNe may be accounted with caution. This was pointed out recently by \cite{2017ApJ...837..120G} in their recent revision of LOSS SN sample, which although it contains targeted SNe, it is volume-limited; and also 
(ii) despite this enhanced number of stripped envelope SNe, when all three subtypes are compared a clear sequence Ic-Ib-IIb from high to low metallicity emerges. Figure \ref{fig:ohNT} shows how the ratio of Ic/Ib is highly dependent on metallicity, and also the upperpanel shows the SN IIb distribution shifted towards lower metallicities compared to SN Ib distribution.


\section{{\sc Hii} region statistics}\label{sec:hii} 

We have seen in Section \ref{sec:local} that differences in the environment of different SN types can be affected by the sample selection (e.g., local oxygen abundance shown in \mbox{Figures \ref{fig:localoh} and \ref{fig:ohNT}}). These local measurements are compared among different galaxies but are not normalized out by any global property. For instance, we may estimate higher local elemental abundance in the environment of a SN exploding in the metal-richest region of a galaxy with low metal abundance than another SN in the metal-poorest environment in a galaxy that is more chemically evolved. 

\begin{figure*}
\centering
\includegraphics[trim=0cm 0cm 0cm 0cm,clip=true,width=\textwidth]{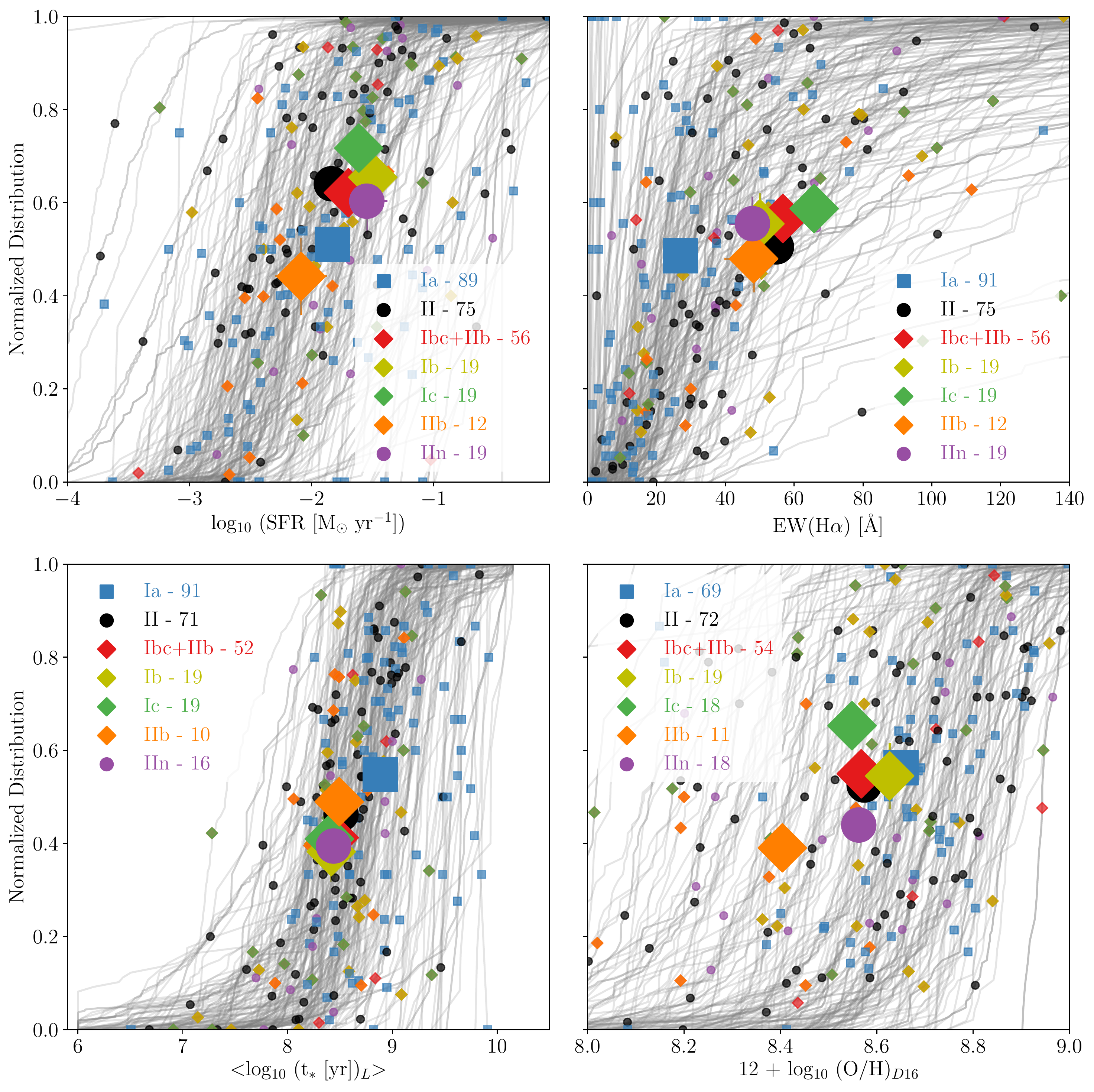}
\caption{{\sc Hii} region cumulative distributions of SFR, H$\alpha$ equivalent width, averaged light-weighted stellar age, and oxygen abundance in all galaxies.
Each grey line represents one galaxy, while small symbols determine the position and the galaxy parameter at the SN parent {\sc Hii} region. 
Seven big symbols, in the foreground, represent the average positions of each SN type.}
\label{fig:hiihaew}
\end{figure*}

Pixel statistics techniques, such as the Normalized Cumulative Rank (NCR) method used, e.g., in \cite{2012MNRAS.424.1372A} and \cite{2018MNRAS.473.1359L} with H$\alpha$ narrow-band imaging as a proxy for SFR, have proved to be useful to find differences among objects when the samples compared are of large enough size.
In \cite{2016MNRAS.455.4087G} we presented a similar approach to the NCR method but using {\sc Hii} regions instead of individual pixels:
we constructed galaxy-wide distributions of the several parameters measured in all {\sc Hii} regions of the galaxy, and positioned the parent SN {\sc Hii} region values in those distributions.
In this way, the locations of SN parent regions in these distributions are already normalized and put into the same scale (0 to 1) allowing useful comparisons among objects occurring in galaxies with a wide range of characteristics.
These distributions contain additional information than the analysis presented in Section \ref{sec:local}. If a SN type was just randomly drawn from the population within galaxies then one expects a y-axis average of 0.5. 
An average rank of a particular SN type that significantly deviates from 0.5 is suggesting that this SN type tends to prefer environments with certain properties {\it within their hosts}.
We remark that to construct these distributions IFS data is essential because it allows the extraction of several {\sc Hii} region integrated spectra from a single datacube, otherwise lots of longslit spectra would be needed to get spectra of all {\sc Hii} regions in a galaxy.

The nearest {\sc Hii} region from the SN position was selected by measuring the deprojected distance from the SN position to the center of all {\sc Hii} regions, and looking for the lowest value. For the deprojection of distances we used the H$\alpha$ velocity map and assumed circular rotation.
Given the short elapse (few tens of Myr) between the birth of CC SN progenitor stars and CC SN explosions, these closest regions were considered to be the parent SN {\sc Hii} regions.
On the other hand, SNe Ia have long delay times (100 Myrs to Gyrs) which make it more problematic to associate any given environment directly to the SN progenitor population, as it could have moved a considerable distance before explosion \citep{2002MNRAS.336..785S}.
However, the analysis of these explosion sites (see e.g., \citealt{2013A&A...560A..66R,2015MNRAS.448..732A,2015Sci...347.1459K}) can still provide insights on progenitor scenarios by direct comparison with SN observed properties. In addition, a big question in SN cosmology is to determine to which extent the environment affects systematic uncertainties when determining extragalactic distances. Several correlations have been found using integrated total measurements of galaxy properties, but local environments can provide more consistent correlations (e.g., \citealt{2013A&A...560A..66R} with $\Sigma_{\rm H\alpha}$;  \citealt{2017arXiv170607697R} with U-V color used as a proxy for age).

Figure \ref{fig:hiihaew} shows the resulting {\sc Hii} region distributions of the four parameters studied in Section \ref{sec:local}: SFR intensity, H$\alpha$ equivalent width, light-weighted average stellar age, and oxygen abundance. Each line (in grey) contains measurements in all {\sc Hii} regions of a single galaxy, and the overploted (small) symbol represents the value of the SN parent region. 
Then, for each SN type we averaged out the measurements at all SN parent {\sc Hii} regions (X-axis), and the position within their own host galaxy normalized distribution (Y-axis).
The resulting averages are shown in larger symbols.
Differences in the X-axis can be directly compared to the results in Section \ref{sec:local}, this time being from measurements performed at the nearest {\sc Hii} region instead of at the SN position.
However, differences in the Y-axis are giving new information: where the SN parent cluster is located within all regions of the same galaxy. In this way, a SN type with higher average values in the Y-axis would mean that they systematically occur at higher values, independently of how this parameter compares between the host and other galaxies.
This information can only be extracted efficiently from wide-field IFU data.
It highlights the usefulness of wide-field IFUs to put the SN parent cluster in context with all other {\sc Hii} regions in the same galaxy.

We have already shown that while new observations of low-mass galaxies have provided a sample of galaxies more similar to those from unbiased searches (see \ref{sec:mass}), most of the SNe in PISCO were discovered by biased surveys, and therefore our CC SN sample might not be representative enough of the whole population. This is for instance confirmed by our {\sc Hii} region analysis of the SFR. 

The upper-left panel of Figure \ref{fig:hiihaew} shows that, consistent with our results from Section \ref{sec:sfh}, SNe Ic, Ib, and IIn parent {\sc Hii} regions have slightly higher SFR on average than SNe Ia and IIb. However, it is now clearer that SNe Ic systematically occur at regions with the highest SFR in their host galaxies, compared to both hydrogen-free CC SNe (Ib/IIb) and to SNe II. 
On the other hand, SNe Ib occur at {\sc Hii} regions with higher SFR than SNe II, but these regions are at the same rank within their host galaxies. 
Again, we find that the most important difference in SFR is between SNe IIb and all other subtypes, the former happening at regions with low SFR and not particularly active within their hosts.

Regarding EW(H$\alpha$), the most significant difference between local and {\sc Hii} region distributions is the lower EW(H$\alpha$) of SNe IIn parent {\sc H ii} regions compared to their locations. All SN type averages are within the 45\% and 60\% rank within their hosts, but SNe Ic parent regions have both larger EW(H$\alpha$), and are slightly above the other SN types in terms of ranking within their own host galaxies.

All CC SNe type parent region averages, except for SNe IIb, lie on similar stellar ages and within regions that are younger on average than all regions in their hosts. SNe Ia occur on average at regions with older ages, and SNe IIb fall in between separated from other CC SNe averages.

For oxygen abundances, we find SN Ia and SN Ib parent regions have on average higher metallicities than other subtypes, but SNe Ic is the type that tends to occur in metal-richer regions within their host galaxies. Again, SNe IIb are significantly separated from all other types at lower metallicity regions both in absolute terms and within their hosts.

In summary, we find SNe Ib occur at locations with higher absolute SFR and higher absolute O/H compared to Ic locations, but SNe Ic are in higher SFR and O/H locations within their host galaxies compared to SNe Ib, thus strengthening the conclusion that SN Ic arise from the highest progenitor metallicity and mass of any CC type analyzed.
We stress here that while any difference in the local values studied in Section \ref{sec:local} may simply just be due to biases in the definition of the sample, when the SN local environment is compared (normalized) to all other environments of the same host galaxy, comparisons among SN types are more fair. This is another demonstration of the power of IFS in studying SN environments.


\section{Implications for core collapse supernova progenitors}\label{sec:prog}

Single-star evolution models predict that the lower zero-age main sequence mass limit for a star to produce a CC SN is around 8 \Msun, which corresponds to a lifetime of $\sim$40 Myr (e.g., \citealt{2000A&AS..141..371G,2003ApJ...591..288H,2009A&A...502..611G}).
Given that the majority of massive stars form in binary systems \citep{2012Sci...337..444S}, more recent theoretical models have considered binaries as an alternative channel for CCSN progenitors. Recently, \cite{2017A&A...601A..29Z} modeled the effects of binary interactions in the delay time distribution (DTD) of CC SNe, and demonstrated that the minimum mass for a massive star to explode as a SN can be as low as 4 \Msun. This lower limit corresponds to longer stellar lifetimes, and the sharp age cutoff for CC SN progenitors at 40 Myr becomes a slow decrease of the DTD, with a tail reaching ages up to $\sim$200 Myr.

Single RSGs have been found at the locations of SNe II in pre-explosion images, and have been undoubtedly associated as SNe II progenitors \citep{2015PASA...32...16S}.
For stripped-envelope supernovae, the picture is much more confusing. 
Their progenitors lose the outer layers by some mechanism before explosion and, under the framework of single-star progenitors, Wolf-Rayet (WR) stars have been historically attributed as their most probable progenitors.
However, as \cite{2011MNRAS.412.1522S} pointed out, classical WR stars could not be the only progenitor channel of SNe Ibc, simply because there are not enough WR stars to reproduce their rates and their fraction within all CC SNe ($\sim$30\%; \citealt{2017ApJ...837..120G}).
This is supported by the fact that all six progenitor detections of SNe Ibc (1 Ib and 5 IIb) were indeed compatible with binary systems (iPTF13bvn,
\citealt{2016MNRAS.461L.117E,2016ApJ...825L..22F};
1993J, \citealt{2004Natur.427..129M};
2008ax, \citealt{2015ApJ...811..147F};
2011dh, \citealt{2014ApJ...793L..22F};
2013df, \citealt{2014AJ....147...37V};
2016gkg, \citealt{2017ApJ...836L..12T}).

Although these findings would rule against single-stars being the exclusive channel, other evidence would favor single-star progenitors. 
For instance, in binary models typically a thin layer of hydrogen is left after the Roche-lobe overflow, given that it is difficult to remove the helium layer only by mass transfer \citep{2017MNRAS.470.3970Y}.
The total amount of helium is also systematically smaller in single-star models than in binary-star models.
Therefore, there may still need to be another mechanism (e.g., metallicity driven winds) to explain the expelling of the remaining He layer in SN Ic progenitors, 
or this may simply imply that single star progenitors are more feasible for SNe Ic (WR stars), while both He-rich subclasses (Ib, and IIb) are more dominated by lower-mass progenitors in binary systems.

This picture is consistent with our findings.
The number ratio Ic/Ib (but also Ic/IIb) increases with metallicity (see Fig. \ref{fig:localparams}), and their stellar populations require younger components, as shown in Figures \ref{fig:ohNT} and \ref{fig:localpercents}.
In this case, higher metallicity would enable the extra mechanism needed to get rid of the helium layer, and younger populations would indicate younger and higher mass progenitors.
SNe Ib, on the other hand, have similar level of on-going star formation at their locations than SNe Ic (see Fig. \ref{fig:localparams}), but significantly less young stellar population components and definitely lower metallicities. 
In the same picture, SNe IIb occurr at lower SFR environments, with a larger amount of older SSP components (and lower young SSPs), and at very low metallicity.
SNe IIb would perfectly fit in the binary system progenitor scenario with two lower mass stars, while SN Ib progenitor systems could be either a lower metallicity single-star case of WR stars \citep{2010A&A...516A.104L,2007ARA&A..45..177C,2005A&A...442..587V}, a binary system of two stars in different stages of evolution and at higher metallicity \citep{2011A&A...528A.131C,2017A&A...601A..29Z}, or a mixture of both scenarios.

Metallicity is clearly playing a role between SN Ic/Ib and SNe Ib/IIb, but other factors (e.g., progenitor mass or age) might be more important in determining the final fate of hydrogen-rich (II) vs. stripped-envelope SNe. 
The correlation with increasingly younger populations in the sequence Ic $\Rightarrow$ Ib $\Rightarrow$ II is usually interpreted within the single star scenario as a difference in progenitor mass and age, but could instead be explained by current binary model predictions \citep{2015PASA...32...15Y}.

Another interesting result from our analysis is the bimodal SFH found at SNe IIn locations. 
SNe IIn are hydrogen-rich CC SNe that suffer some kind of interaction mostly due to circumstellar matter (CSM) in their close environments as indicated by the narrow Balmer emission seen in their spectra. 
This suggests that their progenitors have passed episodes of mass-loss activity prior to explosion, by interaction either with their own stripped material by strong winds, or from a companion star in binary systems.
As suggested by \cite{2015A&A...580A.131T}, if the mass-loss is due to stellar winds we expect to see continuity in the observed properties of SNe IIn, 
however, SNe IIn are the SN type with the most heterogenous observational properties (even though they only account for 5\% of all CC SNe
\citealt{2017ApJ...837..120G}), which may translate in a variety of progenitors. 

There have been at least four progenitor detections of SNe IIn in pre-explosion images 
(1978K, \citealt{1993ApJ...416..167R};
2005gl, \citealt{2009Natur.462..624G};
2009ip, \citealt{2013MNRAS.430.1801M,2013ApJ...767....1P};
2015bh, \citealt{2016MNRAS.463.3894E}),
and in all cases the object found at the SN IIn location was compatible with very massive and young stars, making the transitional luminous blue variables (LBV) the most favored progenitors.  
Although LBVs are usually associated to a transitional  phase between a bright O-star and a WR, \cite{2015MNRAS.447..598S}  examined their locations in the Galaxy and the Magellanic Clouds and proposed that they may be produced in binaries instead.

\cite{2013A&A...555A..10T} showed that SN IIn environments exhibit a metallicity distribution that closely matches that of SNe II, consistently with what \cite{2012MNRAS.424.1372A} found studying the association of SN types to the distribution of the H$\alpha$ emission (proxy for ongoing SFR) in their host galaxies: SN IIn distribution was quite similar to that of SNe II, indicating that they may come from progenitors with similar stellar masses.
\cite{2017A&A...597A..92K} studied the distribution of massive stars and SNe in the LMC and M33 and also found no correlation between LBV and SNe IIn NCR distributions, which favored lower mass progenitors for SNe IIn.
Moreover, \cite{2017arXiv171105765K} recently found that SN IIn is the type less related to ongoing SFR studying the parent clusters with high-resolution narrow-field IFS.
RSG progenitors with superwinds have also been proposed to be progenitors of a few individual SNe IIn (e.g., \citealt{2002ApJ...572..350F,2009AJ....137.3558S}).
An explanation was given in models by \cite{2014Natur.512..282M}, where RSG produces a static shell of CSM confined close to the star, that would produce the narrow lines seen in their spectra. 

\begin{figure*}
\centering
\includegraphics[trim=0cm 0cm 0cm 0cm,clip=true,width=\textwidth]{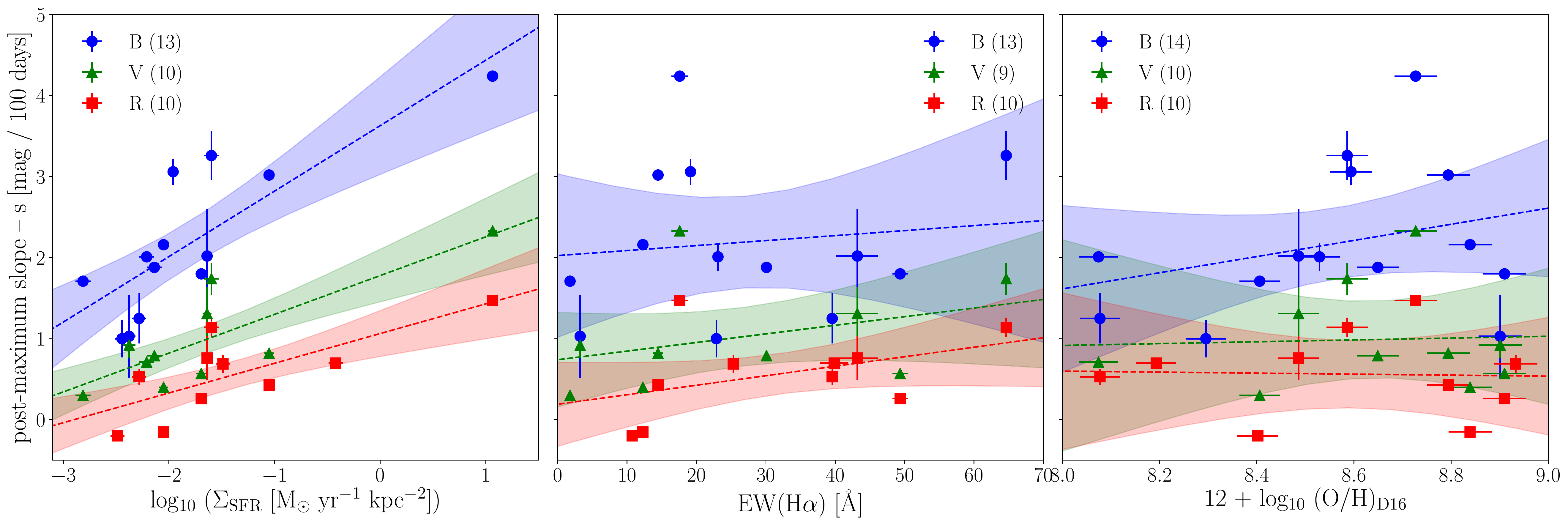}
\caption{Type II SN post-maximum brightness decline rate ($s$) in B (blue), V (green), and R (red) bands as a function of host galaxy parameters measured at SN locations. We found a significant correlation between $s$ in all bands and the local SFR intensity, while no correlation is found with the H$\alpha$ equivalent width and oxygen abundance.}
\label{fig:sniiloc}
\end{figure*}

Our results would explain why detections favored LBVs while SN IIn environments were similar to SN II RSG progenitors: they are very likely to consist of a mixture of very massive and low mass progenitors.
Distributions of SN IIn local environmental parameters (in Fig$.$ \ref{fig:localparams}) showed, on average, higher SFR intensity, younger ages, and higher EW(H$\alpha$) than SNe II. 
This is confirmed when studying their environments within their host galaxies (shown in Fig$.$ \ref{fig:hiihaew}) where similar results are found with respect to SNe II.
But the most significant result is that in our SFH analysis (Fig$.$ \ref{fig:localpercents}) SNe IIn had a bimodal distribution of stellar ages, 
their signal in the innermost and in the oldest stellar population age bin is higher than both the average stripped envelope SNe and SNe II.
Therefore, this would suggest that SNe IIn come from a mixture of progenitors, a fraction compatible with young and massive progenitors (e.g., LBVs), and probably a higher fraction coming from older progenitor populations (e.g., RSGs), both with a presence of shell of hydrogen close-by the star.
Unfortunately, there is no IFS data available for a sample of SNe IIn discovered from unbiased surveys, and we were not able to add such distribution of environmental metallicities in Figure \ref{fig:ohNT}. This will be the topic of a future work, together with a study of the connection SNe IIn and other interacting SN subtypes (Ibn, \citealt{2008MNRAS.389..113P}; Ic/IIn, \citealt{2017arXiv171200027K}; 
Ia/IIn, \citealt{2003Natur.424..651H}).


\section{SN II light-curve and environmental properties linked}\label{sec:sniilocal} 

Hydrogen-rich SNe (type II) have been historically divided in `plateau' (IIP) or `linear' (IIL), depending on the post-maximum brightness decline: type IIP showed a relatively constant brightness for 60-100 days with a sudden decrease plus linear radioactive decay, while type IIL showed a clear decline starting right after maximum light \citep{1979A&A....72..287B}.
It has been previously claimed (e.g., \citealt{2010ApJ...721..777A,2014MNRAS.445..554F}) that hydrogen-rich SNe II formed two distinct classes (IIP/IIL) powered by different mechanisms (e.g., magnetars \citealt{2010ApJ...717..245K}; electron capture; \citealt{1991ApJ...374..266S}). 
However, using larger and more complete samples, other works have shown that the post-maximum decline brightness covers a continuum range of values from totally flat behavior to quite steep, in the V-band \citep{2014ApJ...786...67A} but also in all other optical bands \citep{2015ApJ...799..208S,2015MNRAS.451.2212G, 2016AJ....151...33G, 2016MNRAS.459.3939V}. 
Moreover, \cite{2014ApJ...786...67A} and \cite{2015MNRAS.448.2608V} showed that some SNe II, which would have been classified as a typical SN IIL, had this subtle decrease in brightness at the expected epoch (60-100 days) of other SNe IIP.
Therefore, hydrogen-rich SNe are starting to be classified as fast- or slow-decliners using different parameterizations (e.g., decline slope after transition from cooling to recombination phases, s$_2$, or after peak brightness, $s$, until the end of recombination phase) and break points.
Given that this family spans a continuum in several observed and physical properties, some continuous parameter(s) has(ve) to be responsible for such a sequence.

\begin{deluxetable}{lccc}
\tablenum{5}
\tablecolumns{4}
\tablewidth{0pt}
\tablecaption{Post-maximum brightness declines in BVR bands for the 17 SNe used in Section \ref{sec:sniilocal} and Figure \ref{fig:sniiloc}. \label{tab:sniiloc}}
\tablehead{
\colhead{\textbf{SN}} & \colhead{\textbf{s$_B$}} & \colhead{\textbf{s$_V$}} & \colhead{\textbf{s$_R$}}  \\
 & \colhead{[mag / 100 days]} & \colhead{[mag / 100 days]} & \colhead{[mag / 100 days]}} 
\startdata
\cline{1-4}    
 PTF09cjq  &    $---$    &     $---$    &  0.69 (0.11) \\
 SN 1948B  & 2.01 (0.17) &     $---$    &     $---$    \\
 SN 1994N  & 1.00 (0.23) &     $---$    &     $---$    \\
 SN 1999bg & 2.01 (0.04) &  0.71 (0.03) &     $---$    \\
 SN 1999em & 2.16 (0.01) &  0.40 (0.01) & -0.15 (0.01) \\
 SN 1999gi & 1.80 (0.03) &  0.57 (0.03) &  0.26 (0.04) \\
 SN 2003gd & 2.02 (0.58) &  1.31 (0.33) &  0.76 (0.27) \\
 SN 2003hg & 3.02 (0.04) &  0.82 (0.02) &  0.43 (0.02) \\
 SN 2005au & 3.26 (0.30) &  1.74 (0.20) &  1.14 (0.12) \\
 SN 2005dp & 1.25 (0.31) &     $---$    &  0.53 (0.10) \\
 SN 2006be & 1.88 (0.04) &  0.79 (0.01) &     $---$    \\
 SN 2006ee & 1.71 (0.04) &  0.30 (0.03) &     $---$    \\
 SN 2007Q  & 1.03 (0.51) &  0.92 (0.16) &     $---$    \\
 SN 2008ij & 3.06 (0.16) &     $---$    &     $---$    \\
 SN 2011cl &    $---$    &     $---$    &  0.70 (0.04) \\
 SN 2013cf &    $---$    &     $---$    & -0.20 (0.05) \\
 SN 2013ej & 4.24 (0.06) &  2.23 (0.03) &  1.49 (0.02) \\
\enddata
\end{deluxetable}

With this aim, we looked at correlations between SFR intensity, oxygen abundance, and H$\alpha$ equivalent width measured at the SN II locations and BVR broad band light-curve parametrization defined in \cite{2014ApJ...786...67A}, for those objects in PISCO with available light-curves to check if any environmental parameters might be behind that sequence.
In Figure \ref{fig:sniiloc} we show how the {\it post-maximum} brightness decline ($s$) relates to these local parameters, and all relevant numbers are presented in Table \ref{tab:sniiloc}.

Only the correlation between $s$ and the $\Sigma_{\rm SFR}$ is significant in all three bands according to a Pearson test (factors of 0.82, 0.81, and 0.73 in the B, V, and R band, respectively)\footnote{The relation is not driven by the point at high $\Sigma_{\rm SFR}$ as shown by the still high correlation factors found when this point is not considered: 0.69, 0.49, 0.50, respectively for BVR bands.}.
Lower correlation factors (from 0.1 in B to 0.4 in R) were found for H$\alpha$ equivalent width, and no correlation ($\sim$0.0 in V and R bands, and 0.3 in the B band) was found for O/H, in agreement with results presented in \cite{2016A&A...589A.110A}. 
When using the brightness decline parameter {\it during the plateau phase} ($s_2$), while the significance on the $\Sigma_{\rm SFR}$  relation decreases to \mbox{(B, V, R)=(0.50, 0.75, 0.67)}, the relation with EW(H$_\alpha$) increases its significance to \mbox{(B, V, R)=(0.56, 0.34, 0.49)}\footnote{While $s$ accounts for the brightness decline rate from the epoch maximum light to the end of the plateau phase, $s_2$ takes into account that in some SNe II the transition between the cooling and the recombination phase becomes evident as a variation (slowdown) of the brightness decline. While $s$ contains information from both phases, $s_2$ purely relates to the amount of hydrogen in the outer layer.}.

The dependence of the post-maximum brightness decline on the local SFR intensity has never been studied before, and may be key in explaining the diversity of the slopes in SNe II: fast-declining SNe II, traditionally labeled as IIL, would occur at regions with denser SFR compared to slow-declining SNe II, traditionally labeled as IIP.
Our results would disfavour different channels for different SNe II, and instead differences may be explained by other continuous properties proxied by $\Sigma_{\rm SFR}$.

Given that the SFR was measured using the local H$\alpha$ flux as a proxy, and therefore it is tracing very recent and ongoing SF (up to a few tens of Myr), this correlation may also indicate a sequence in progenitor age lifetimes, in the sense that fast-decliners would come from younger progenitors than slow-decliners. 
This is in line with the result from \cite{2012MNRAS.424.1372A}, where SN IIL locations were shown to be more correlated to the distribution of SF in their hosts than SNe IIP, as shown in their NCR diagram constructed from H$\alpha$ narrow-band imaging. 

However, given the low number of objects used here, this trend deserves a further confirmation with more data. This is the goal of an ongoing specific work combining PISCO with other IFS data, and will be presented elsewhere.


\section{Summary and conclusions}\label{sec:conc} 

In this paper we presented the PMAS/PPak Integral-field Supernova hosts COmpilation (PISCO), an ongoing effort to complete the SN host galaxy sample with IFU observations selected from the CALIFA survey including low-mass CC SN hosts and objects with SN Ia light-curves available.
These observations cover the whole galaxy extent including the SN local environment, and they are all observed with the same instrumental configuration. 
PISCO consists of 466,347 spectra of 232 host galaxies that hosted 272 SNe of all types, with an average spatial resolution of 380 pc, and where we were able to dectect 11,270 {\sc Hii} regions.

While the SN Ia host galaxy sample in PISCO is statistically consistent in stellar mass to SN Ia hosts from untargeted surveys, CC SNe hosts in PISCO are still slightly biased towards high-mass galaxies, characteristic from targeted searches. However this bias has been reduced compared to the CC SN host sample presented in \cite{2016AJ....151...33G}.
We constructed distributions of SFR intensity, EW(H$\alpha$), light-weighted average stellar age, and O/H measured at SN locations and recovered previous sequences in terms of association of these parameters with different SN types.
We reconstructed the SFHs at all SN locations dividing the contributions of young SSPs in 4 bins between 0 and 300 Myr.
For the first time, {\sc Hii}-region distributions of several parameters were constructed for each galaxy, in order to study how the SN parent region was ranked within {\sc Hii} regions of the same galaxy. Then, we averaged out all parameters and ranks for each SN type.
Our main results are:

\begin{itemize}

\item We found that SNe Ic occur at higher SFR, higher EW(H$\alpha$), and more metal-rich {\sc Hii} regions within their hosts, as compared to the mean values within those same galaxies.
SN Ic environments need a larger fraction of young (0-40 Myr) SSP models to reproduce their spectra than other CC SN types which, together with the highest association of this SN type to $\Sigma_{\rm SFR}$ environments and the higher EW(H$\alpha$) found at their locations, supports the picture of having more massive progenitors than other CC SNe.

\item SNe IIb environments are the most different compared to other CC SN types, both in general and within their host galaxies. SNe IIb occur in metal-poor, low EW(H$\alpha$), and with relatively low SFR environments. They also have the lower component of young SSP in the innermost bin of the SFH at their locations, and the highest (with SNe IIn) in the oldest bin. 
They would perfectly fit in the binary system progenitor scenario of two lower mass stars.

\item SNe Ib fall in between SNe Ic and SNe IIb in all our diagrams. We suggest that both scenarios, single stars with weaker stellar winds as compared to SNe Ic (given that the SNe Ic/Ib increases with metallicity), and binary stars with stronger mass transfer than for SNe IIb, may be at play.

\item We detect an excess of both very young and old SSPs in the SFHs of SNe IIn environments compared to other SN types. This would support the picture of different progenitors for both components: LBVs for the young component, and RSGs for the old component.

\item We present for the first time a correlation between observed light-curve parameters and local environmental properties of SNe II. In particular, we found that the SN II post-maximum brightness decline in all three BVR bands shows a trend with the local SFR intensity, which could explain the range of decline rates of fast-declining (IIL) and slow-declining (IIP) SNe II, and indicate a difference in progenitor age lifetimes: lower ages and therefore higher mass for fast declining SN~IIL as compared to slower declining SNe~IIP.

\item Our results highlight the need of constructing samples of IFS observations built from unbiased survey(s) of SNe, also volume-limited, to get reliable constrains on associations of different SN types to environmental properties. 
\end{itemize}

PISCO reduced datacubes and dataproducts generated during the analysis, such as emission-line maps, {\sc H ii} region segregation maps and {\sc H ii} region spectra, are made publicly available from the 
PISCO github repo\footnote{\href{https://github.com/lgalbany/pisco}{https://github.com/lgalbany/pisco}}. With this paper we release all CC SN host galaxies, while current SN Ia hosts dataset will be published with a forthcoming paper. PISCO is an ongoing project focused on different aspects of supernova science, and future publications will be accompanied with the release of the specific dataset used with that goal.
%

\section{Future work} \label{sec:future}

One important caveat in the distinction of objects discovered by targeted/untargeted surveys is that, although untargeted surveys discover transients unbiasedly in terms of galaxy properties, given the large amount of objects detected and the limited resources for classification and follow up, there could still be some bias in the object prioritization. 
Therefore, it is important to construct a SN sample that is both unbiased in terms of discovery but also with a clear constraint on follow up priority.
We anticipate that in the near future, in order to improve current environmental analyses and to alleviate biases such as the host galaxy mass shown in Section \ref{sec:mass},
it would be beneficial to define homogeneous samples of SNe discovered blindly.
The Large Synoptic Survey Telescope (LSST; \citealt{2017arXiv170804058L}) and the Zwicky Transient Factory (ZTF; \citealt{2015AAS...22532804B}) open an opportunity as feeder surveys for an IFU blind follow-up program for both CC SNe and SNe Ia which, in the case of SNe Ia, would be an excellent low-z companion survey for Wide-Field InfrarRed Survey Telescope (WFIRST; \citealt{2015arXiv150303757S}) SN survey\footnote{Assuming that the Integral Field Unit Channel (IFC) is included in the final design of the instrument.}.

To improve spatial resolution, we have started a new effort, the All-weather MUse Supernova Integral-field Nearby Galaxies (AMUSING; \citealt{2016MNRAS.455.4087G}) that takes advantage of the combination of the wide-aperture 8.1m Very Large Telescope (VLT) at Cerro Paranal, and the Multi-Unit Spectroscopic Explorer (MUSE; \citealt{2014Msngr.157...13B}) Integral Field Unit, which provides with a large FoV (1'$\times$1') and an impressive spatial resolution (0.2" $\times$ 0.2" per spaxel) only limited by seeing. With almost 100,000 spectra from $\sim$4650-9000~\AA~and R$\sim$3500-1800 (from the blue to the red extent) delivered per pointing, the field of supernova environments has entered into the regime of big data. Currently, after 5 semesters collecting data, the AMUSING sample consists of more than 300 nearby SN host galaxies, it has already delivered a few analyses on both galaxy enrichment and evolution \citep{2015A&A...573A.105S,2016ApJ...830L..40S,2017arXiv171001188S,2017arXiv171102785L} and supernova environments \citep{2016MNRAS.455.4087G,2016ApJ...830L..32P,2017A&A...602A..85K,2016A&A...593A..78K,2017arXiv171105765K}, and will certainly revolutionize the field in the coming years.

\acknowledgments

We are greatly in debt with the Calar Alto Observatory staff because without their kindness and interest for this project, most of the observations would have not been possible. 
We are greatly thankful to Yao-Yuan Mao for being behind most of the Python coding developed for plotting our results, as well as Or Graur for useful discussions.
L.G. and W.M.W-V.  were supported in part by the US National Science Foundation under Grant AST-1311862.
I.D. is funded by the MINECO-FEDER AYA2015-63588-P grant.
Escrit en la seva major part a cavall de la Biblioteca de Palafrugell i la Platgeta de Calella.

Based on observations collected at the Centro Astron\'omico Hispano Alem\'an (CAHA) at Calar Alto, operated jointly by the Max-Planck-Institut f\"ur Astronomie (MPIA) and the Instituto de Astrof\'isica de Andaluc\'ia (CSIC).
This study makes use of the data provided by the Calar Alto Legacy Integral Field Area (CALIFA) survey (\href{http://califa.caha.es}{http://www.caha.es/CALIFA/}). 
The {\tt STARLIGHT} project is supported by the Brazilian agencies CNPq, CAPES and FAPESP and by the France-Brazil CAPES/Cofecub program.
This research has made use of the NASA/IPAC Extragalactic Database (NED), which is operated by the Jet Propulsion Laboratory, California Institute of Technology, under contract with the National Aeronautics and Space Administration, and data products from SDSS and PanSTARRS surveys. 
Funding for the Sloan Digital Sky Survey IV has been provided by the Alfred P. Sloan Foundation, the U.S. Department of Energy Office of Science, and the Participating Institutions. SDSS-IV acknowledges
support and resources from the Center for High-Performance Computing at the University of Utah. The SDSS web site is www.sdss.org.
SDSS-IV is managed by the Astrophysical Research Consortium for the  Participating Institutions of the SDSS Collaboration including the  Brazilian Participation Group, the Carnegie Institution for Science,  Carnegie Mellon University, the Chilean Participation Group, the French Participation Group, Harvard-Smithsonian Center for Astrophysics, Instituto de Astrof\'isica de Canarias, The Johns Hopkins University, Kavli Institute for the Physics and Mathematics of the Universe (IPMU) / University of Tokyo, Lawrence Berkeley National Laboratory, Leibniz Institut f\"ur Astrophysik Potsdam (AIP),  Max-Planck-Institut f\"ur Astronomie (MPIA Heidelberg), Max-Planck-Institut f\"ur Astrophysik (MPA Garching), Max-Planck-Institut f\"ur Extraterrestrische Physik (MPE), National Astronomical Observatories of China, New Mexico State University, New York University, University of Notre Dame, Observat\'ario Nacional / MCTI, The Ohio State University, Pennsylvania State University, Shanghai Astronomical Observatory, United Kingdom Participation Group,Universidad Nacional Aut\'onoma de M\'exico, University of Arizona, University of Colorado Boulder, University of Oxford, University of Portsmouth, University of Utah, University of Virginia, University of Washington, University of Wisconsin, Vanderbilt University, and Yale University.
The Pan-STARRS1 Surveys (PS1) and the PS1 public science archive have been made possible through contributions by the Institute for Astronomy, the University of Hawaii, the Pan-STARRS Project Office, the Max-Planck Society and its participating institutes, the Max Planck Institute for Astronomy, Heidelberg and the Max Planck Institute for Extraterrestrial Physics, Garching, The Johns Hopkins University, Durham University, the University of Edinburgh, the Queen's University Belfast, the Harvard-Smithsonian Center for Astrophysics, the Las Cumbres Observatory Global Telescope Network Incorporated, the National Central University of Taiwan, the Space Telescope Science Institute, the National Aeronautics and Space Administration under Grant No. NNX08AR22G issued through the Planetary Science Division of the NASA Science Mission Directorate, the National Science Foundation Grant No. AST-1238877, the University of Maryland, Eotvos Lorand University (ELTE), the Los Alamos National Laboratory, and the Gordon and Betty Moore Foundation.

\facility{CAO:3.5m(PMAS/PPak).}
\software{Numpy \citep{2011arXiv1102.1523V}, Matplotlib \citep{2007CSE.....9...90H}, STARLIGHT \citep{2005MNRAS.358..363C}, HIIexplorer  \citep{2012A&A...546A...2S}.}

\bibliographystyle{yahapj}
\bibliography{pisco}

\appendix

\section{Description of previous observations} \label{app:a}

Around half of the observations in PISCO have been collected from the following sources:

\begin{itemize}
\item[(i)] 115 galaxies that hosted 128 SNe were either included in the CALIFA 3rd data release or observed under a few parallel programs led by CALIFA members that were labeled together as {\it CALIFA-extensions}. More details about these CALIFA-extensions are described in the CALIFA 3rd data release \citep{2016A&A...594A..36S}. 

\item[(ii)] Two galaxies, NGC 5668 \citep{2012ApJ...754...61M} and NGC 3982, were observed under two programs that aimed to study these two particular objects (PI: Marino). They hosted 3 SNe in total.

\item[(iii)] Four galaxies, NGC 0105, UGC 04008, CGCG 207-042, and UGC 05129, were included in a pilot project to study SN Ia host galaxies with Integral Field Spectroscopy \citep{S12}. 

\item[(iv)] Eight objects are observations from PPak IFS Nearby Galaxies Survey \citep[PINGS; ][]{2010MNRAS.405..735R}. These include a few large ($>$12 pointing) mosaics of wide-field nearby galaxies (e.g., NGC 0628)
\end{itemize}

Subsamples of galaxies (81, 115, and 23, respectively) from the above sources were used in \cite{2014A&A...572A..38G,2016A&A...591A..48G,2017MNRAS.468..628G}, where the properties of the SN local environment in terms of star-formation, metallicity and extinction, respectively, were studied.

\section{Description of new observations} \label{app:b}

In Table \ref{tab:prop2} we list the details of all nights dedicated to each of the 5 programs in PISCO indicating the success ratio. Note that due to unfortunate bad weather nights, the observatory was usually able to award additional nights at the end of the semester that contributed to finish or increase the success ratio of the observations significantly. Below, we give more details about each program:

{\bf 15B:} We proposed to observe a sample of galaxies that hosted CC SNe with  precise spectral type and that occurred in low-mass galaxies. This has allowed us both to partially correct the bias in the host galaxy mass present in the CALIFA SN host galaxy sample, and to split the CC SN group into different SN subtypes. This paper presented the results of this particular program.

{\bf 16A:} SNe Ia are used as accurate distance indicators, however the nature of their progenitors is still unclear. Differentiating between the single degenerate (WD and non-degenerate donor star) and double degenerate (two WDs) scenarios is a key issue. 
The detection of blueshifted Na D absorption in some SN Ia spectra, and its potential association with circumstellar material structures close to the SN progenitor, has been the focus of recent attention.
In semester 16A we proposed the observation of 30 galaxies (of which 21 were actually observed) that hosted SNe with deep NaD features detected in their spectra, to constrain the nature of the NaD absorption, SN progenitor age and metallicity.
We aimed to understand the origin and nature of narrow NaD absorption in mid-resolution SN Ia spectra, and its relation to either ISM or CSM material  through PMAS observations of their host galaxies.
The sample were selected from works reporting the presence of strong Na D features in SN spectra (basically \citealt{2011Sci...333..856S,2013MNRAS.436..222M}, and \citealt{2013ApJ...779...38P}).

{\bf 16B/17A/17B:}
Although optical observations of SNe Ia have proved essential to measure accurate cosmological distances, they are superior standard candles in the near-infrared (NIR), both because their light curves are intrinsically more similar at these wavelengths, and reddening effects are greatly reduced \citep{2004AJ....128.3034K,2008ApJ...689..377W}. 
SN Ia Hubble residuals using optical data correlate with global host galaxy parameters (such as total mass), and the addition of a term in the SN Ia light curve standardization accounting for these environmental parameters, has proved to reduce further the scatter in the SN Ia absolute magnitude at peak.
We proposed observations of galaxies that hosted SNe Ia that were included in the SweetSpot Survey, a three-year NOAO Survey program that obtained near-infrared ($JHK$) observations of 114 nearby (0.02 $<$ z $<$ 0.09) SNe Ia with the WIYN High-resolution Infrared Camera (WHIRC) on the WIYN 3.5-m telescope \citep{2014ApJ...784..105W,2017arXiv170302402W}. Thirty-four of the 49 proposed observations (9/16, 12/20, and 13/13, per semester respectively) were successfully performed.
Our aim using PISCO data is to complement IFS observations from HexPak \citep{pittir32578} and MUSE \citep{2016MNRAS.455.4087G}, and reduce further the scatter in the NIR SN Ia Hubble diagram by looking for the first time for correlations between SN Ia residuals and both global and local galactic properties. 

\begin{deluxetable}{ccccccc}
\tablecaption{Summary of the time scheduled and executed for each of the observing programs listed in Table \ref{tab:prop}. Last column tracks the completion percent of the initial program. \label{tab:prop2}}
\tabletypesize{\scriptsize}
\tablenum{6}
\tablecolumns{6}
\tablewidth{0pt}
\tablehead{
\colhead{Night} & \colhead{Observed objects} & frac. night & success &scheduled&completion\\
{\tiny(AAMMDD)} &  &  &  &  & }
\startdata
\cline{1-6}              
\multicolumn{6}{c}{{H15-3.5-004}} \\
\cline{1-6}    
150810 & 4 & 1.00 & 1.00 & 1.00 &  8\% \\
150811 & 0 & 1.00 & 0.00 & 1.00 &  8\% \\
151102 & 0 & 1.00 & 0.00 & 1.00 &  8\% \\
151103 & 1 & 1.00 & 0.09 & 1.09 & 10\% \\
151218 &10 & 1.00 & 1.00 & 1.09 & 30\% \\
151219 &10 & 1.00 & 1.00 & 2.09 & 50\% \\
151220 &10 & 1.00 & 0.91 & 3.00 & 70\% \\
151221 &10 & 1.00 & 1.00 & 4.00  & {\bf 90\%} \\
\cline{1-6}              
\multicolumn{6}{c}{{F16-3.5-006}} \\
\cline{1-6}  
160118 & 1 & 1.00 & 0.23 & 0.23 &  3\% \\
160119 & 0 & 1.00 & 0.00 & 0.23 &  3\% \\
160201 & 3 & 0.45 & 0.45 & 0.68 & 13\% \\
160314 & 4 & 0.75 & 0.67 & 1.33 & 27\% \\
160316 & 4 & 0.91 & 0.60 & 1.94 & 40\% \\
160606 & 1 & 0.50 & 0.50 & 2.44 & 43\% \\
160607 & 5 & 1.00 & 1.00 & 3.44 & 60\% \\
160608 & 3 & 0.75 & 0.75 & 4.19 & {\bf 70\%} \\
\cline{1-6}              
\multicolumn{6}{c}{{H16-3.5-012}} \\
\cline{1-6}  
161008 & 2 & 0.17 & 0.05 & 0.05 & 13\% \\
161009 & 7 & 1.00 & 1.00 & 1.05 & {\bf 56\%} \\
\cline{1-6}              
\multicolumn{6}{c}{{F17-3.5-001}} \\
\cline{1-6}  
170428 & 0 & 1.00 & 0.00 & 0.00 &  0\% \\
170429 & 0 & 1.00 & 0.00 & 0.00 &  0\% \\
170430 & 5 & 1.00 & 1.00 & 1.00 & 25\% \\
170525 & 3 & 0.45 & 0.45 & 1.45 & 40\% \\
170526 & 4 & 0.58 & 0.58 & 2.03 & {\bf 60\%} \\
\cline{1-6}              
\multicolumn{6}{c}{{H17-3.5-001}} \\
\cline{1-6}  
170817 & 6 & 1.00 & 1.00 & 1.00 &  50\% \\
170818 & 7 & 1.00 & 1.00 & 2.00 &  {\bf 100\%} \\
\enddata
\end{deluxetable}

\begin{deluxetable*}{lclcccccccr}
\tabletypesize{\scriptsize}
\tablenum{2}
\tablecaption{Properties of the 232 galaxies and 272 SNe included in PISCO (semesters 15B to 17B). } \label{tab:sam}
\tablehead{
\colhead{\textbf{Galaxy name}} 
& \colhead{\textbf{ID}}
& \colhead{\textbf{Morphology}} & \colhead{\textbf{Galaxy RA}} & \colhead{\textbf{Galaxy Dec}} & \colhead{\textbf{z}} & \colhead{\textbf{SN}} & \colhead{\textbf{Type}} & \colhead{\textbf{SN RA}} & \colhead{\textbf{SN Dec}} & \colhead{\textbf{Sep.}}
}
\startdata
\cline{1-11}              
\multicolumn{11}{c}{{From CALIFA DR3}} \\
\cline{1-11}              
UGC 00005					&    2	& SABbc      			& 00:03:05.66	&$-$01:54:49.7	& 0.024253	& 2000da				& II			& 00:03:06.52	& $-$01:54:41.8	&  15.1 \\
							&     	&            			&            	&             	& 			& 2003lq				& Ia			& 00:03:04.02	& $-$01:54:45.5	&  24.9 \\
							&     	&            			&            	&             	& 			& 2016eob				& II			&  00:03:07.12 & $-$01:54:42.0	&   23.2   \\
UGC 00139					&   11	& SAB(s)c?   			& 00:14:31.85	&$-$00:44:15.2	& 0.013219	& 1998dk				& Ia			& 00:14:32.16	& $-$00:44:10.9	&   6.3 \\
UGC 00148					&   12	& S?         			& 00:15:51.28	&  +16:05:23.2	& 0.014053	& 2003ld				& II			& 00:15:51.85	&   +16:05:21.6	&   8.4 \\
NGC 0214					&   28	& SAB(r)c    			& 00:41:28.03	&  +25:29:58.0	& 0.015134	& 2005db				& IIn			& 00:41:26.79	&   +25:29:51.6	&  18.0 \\
NGC 0309					&   34	& SAB(r)c    			& 00:56:42.66	&$-$09:54:49.9	& 0.018886	& 1999ge				& II			& 00:56:43.76	& $-$09:54:43.0	&  17.7 \\
NGC 0523					&   48	& pec        			& 01:25:20.73	&  +34:01:29.8	& 0.015871	& 2001en				& Ia			& 01:25:22.90	&   +34:01:30.5	&  27.0 \\
NGC 0716                    &   65  & SBa?                  & 01:52:59.68   &  +12:42:30.5  & 0.015204  & 2017fqo               & II               & 01:53:01.51   &   +12:42:46.0 &  30.9     \\
...\\
\enddata
\tablecomments{Table 1 is published in its entirety in the machine-readable format. A portion is shown here for guidance regarding its form and content.}
\end{deluxetable*}

\begin{deluxetable*}{llcccccccccc}
\tablecaption{Milky Way and host galaxy extinction-corrected fluxes and environment parameters measured in the SN local 1~kpc diameter spectra, together with the host galaxy stellar mass.  \label{tab:measurements}
}
\tablenum{3}
\tablewidth{0pt}
\tabletypesize{\tiny}
\tablehead{
\colhead{\textbf{SN name}} & \colhead{\textbf{SN type}} & \colhead{\textbf{F(H$\alpha$)}} & \colhead{\textbf{F(H$\beta$)}} & \colhead{\textbf{F(O III)}} & \colhead{\textbf{F(N II)}} & \colhead{\textbf{F(S II)}} & \colhead{\textbf{log$_{10}$ $\Sigma_{\rm SFR}$}} & \colhead{\textbf{EW(H$\alpha$)}} & \colhead{\textbf{<log$_{10}$ (t$_* [yr])_{\rm L}$>}}  & \colhead{\textbf{12+ log$_{10}$ (O/H)$_{\rm D16}$}} & \colhead{\textbf{log$_{10}$ (M$_*$ [M$_\sun$])}} \\
 & & \multicolumn{5}{c}{[10$^{-16}$ erg s$^{-1}$ cm$^{-2}$ \AA$^{-1}$]} & [dex] & [\AA] & [dex] & [dex] & [dex] 
}
\startdata
\hline
2000da & II & 28.25 (2.00) & 9.88 (0.68) & 1.84 (0.14) & 9.67 (0.70) & 8.25 (0.36) & -1.5234 (0.0319) & 32.25 (0.50) & 8.40 (0.14) & 8.72 (0.02) & 11.17 \\ 
2003lq & Ia & 6.86 (0.50) & 2.40 (0.19) & 1.02 (0.10) & 2.78 (0.20) & 3.42 (0.18) & -2.1378 (0.0326) & 24.39 (0.90) & 8.88 (0.06) & 8.58 (0.02) & 11.17 \\ 
2016eob & II & 6.46 (0.47) & 2.26 (0.15) & 0.84 (0.08) & 2.54 (0.17) & 2.71 (0.11) & -2.1642 (0.0325) & 24.46 (0.58) & 8.97 (0.12) & 8.64 (0.02) & 11.17 \\ 
1998dk & Ia & 80.88 (5.88) & 28.28 (2.09) & 14.35 (1.04) & 24.64 (1.82) & 29.30 (1.52) & -1.6059 (0.0327) & 34.43 (0.32) & 8.26 (0.12) & 8.56 (0.02) & 10.20 \\ 
2003ld & II & 204.00 (14.67) & 71.34 (5.44) & 50.32 (3.67) & 61.00 (4.37) & 74.04 (3.91) & -1.1499 (0.0324) & 44.31 (0.33) & 8.21 (0.15) & 8.55 (0.02) & 10.43 \\ 
2005db & IIn & 159.60 (11.66) & 55.79 (4.06) & 7.17 (0.62) & 53.50 (3.86) & 35.14 (1.84) & -1.1911 (0.0330) & 50.81 (0.50) & 7.90 (0.13) & 8.83 (0.03) & 11.16 \\ 
1999ge & II & 14.82 (1.14) & 5.18 (0.41) & $---$ & 5.67 (0.46) & 3.21 (0.19) & -2.0266 (0.0347) & 17.24 (0.36) & 8.92 (0.10) & 8.91 (0.04) & 11.01 \\ 
2001en & Ia & 3.37 (0.24) & 1.66 (0.10) & 1.24 (0.10) & 1.38 (0.10) & 2.01 (0.09) & -2.8244 (0.0327) & 5.40 (0.42) & 9.10 (0.14) & 8.50 (0.01) & 10.94 \\ 
2017fqo & II & 14.71 (1.23) & 5.34 (0.36) & 6.14 (0.51) & 5.94 (0.35) & $---$ & -2.2223 (0.0380) & 20.00 (1.45) & 8.36 (0.14) & $---$ & 10.77 \\ 
...\\
\enddata
\tablecomments{Table 1 is published in its entirety in the machine-readable format. A portion is shown here for guidance regarding its form and content.}
\end{deluxetable*}

\end{document}